\documentclass{article}

\usepackage{arxiv}
\usepackage{amsmath}
\usepackage{hyperref}       
\usepackage{url}            
\usepackage{booktabs}       
\usepackage{amsfonts}       
\usepackage{nicefrac}       
\usepackage{microtype}      
\usepackage{cleveref}       
\usepackage{graphicx}
\usepackage{cite}
\usepackage{doi}
\usepackage{multirow}
\usepackage{caption}
\usepackage{textcomp}
\usepackage{authblk}
\usepackage{graphicx}

\usepackage{siunitx}    
\usepackage{tabularx}
\usepackage[acronym,nomain,nonumberlist]{glossaries}
\usepackage{glossary-inline}
\usepackage{orcidlink}
\usepackage{eurosym}

\newcommand*{\T}[1]{\ensuremath{\mathrm{T}_{#1}}}
\newcommand*{\B}[1]{\ensuremath{\mathrm{B}_{#1}}}

\makenoidxglossaries
\newacronym{emi}{EMI}{Electromagnetic Interference}
\newacronym{dsv}{DSV}{Diameter Spherical Volume}
\newacronym{gpa}{GPA}{Gradient Power Amplifier}
\newacronym{rfpa}{RFPA}{RF Power Amplifier}
\newacronym{scu}{SCU}{System Control Unit}
\newacronym{txrx}{Tx/RX}{Transmit/Receive}
\newacronym{lna}{LNA}{Low Noise Amplifier}
\newacronym{inrim}{INRiM}{Instituto Nazionale di Ricerca Metrologica}
\newacronym{mdr}{MDR}{Medical Device Regulation}

\newacronym{fid}{FID}{Free Induction Decay}
\newacronym{adc}{ADC}{Analog Digital Converter}
\newacronym{fov}{FOV}{Field of View}
\newacronym{etl}{ETL}{Echo Train Length}
\newacronym{bw}{BW}{Bandwidth}
\newacronym{tse}{TSE}{Turbo Spin-Echo}
\newacronym{t1w}{T1w}{T1-weighted}
\newacronym{pdw}{PDw}{PD-weighted}

\makenoidxglossaries
\glsdisablehyper

\DeclareSIUnit{\ppm}{ppm}
\DeclareSIUnit{\dBm}{dBm}

\title{A Reference System for Open Source Portable Low-Field MRI}

\begin{document}


\author[1]{David Schote\thanks{
    David Schote and Helge Herthum contributed equally to this study. Corresponding author: David Schote, Physikalisch-Technische Bundesanstalt (PTB), Abbestr.\ 2-12, 10587 Berlin, Germany. \href{mailto:david.schote@ptb.de}{david.schote@ptb.de}}\orcidlink{0000-0003-3468-0676}
}
\author[2]{Helge Herthum\textsuperscript{*}\orcidlink{0000-0001-6494-0833}}
\author[3]{Umberto Zanovello\orcidlink{0000-0001-6415-9967}}
\author[4]{Julia Pfitzer\orcidlink{0009-0000-3758-209X}}
\author[2,5]{Ivo Jutte\orcidlink{0009-0000-2348-224X}}
\author[1]{Jan Gregor Frintz}
\author[1]{Ilia Kulikov\orcidlink{0000-0002-7544-2124}}
\author[6]{Sebastian Littin\orcidlink{0000-0002-8190-1761}}
\author[1]{Frank Seifert\orcidlink{0000-0002-7065-2528}}
\author[1]{Christoph Kolbitsch\orcidlink{0000-0002-4355-8368}}
\author[1,7]{Lukas Winter\orcidlink{0000-0002-4381-275X}}

\affil[1]{Physikalisch-Technische Bundesanstalt (PTB) Braunschweig and Berlin, Berlin, Germany}
\affil[2]{Berlin Center for Advanced Neuroimaging, Charit\'e -- Universit\"atsmedizin Berlin, Corporate Member of Freie Universit\"at Berlin, Humboldt-Universit\"at zu Berlin, and Berlin Institute of Health, Berlin, Germany}
\affil[3]{Instituto Nazionale di Ricerca Metrologica (INRiM), Turino, Italy}
\affil[4]{Institute of Biomedical Imaging, Graz University of Technology, Graz, Austria}
\affil[5]{Department of Biomedical Engineering, Eindhoven University of Technology, Eindhoven, The Netherlands}
\affil[6]{Division of Medical Physics, Department of Radiology, Medical Center -- University of Freiburg, Faculty of Medicine -- University of Freiburg, Freiburg, Germany}
\affil[7]{Open Source Imaging Initiative (OSI\textsuperscript{2}) e.V., Berlin, Germany}
 
\date{}

\maketitle

\begin{abstract}
    Portable low-field MRI systems received renewed attention due to their promising applications in point-of-care imaging. However, the pathway to a functional MRI system is challenging which hinders effective adaption across research groups. Due to incomplete documentation, reproducibility is limited and systems are often redesigned, making comparisons across systems difficult. Additionally, system tests and characterizations lack standardization which complicates ethical approval for clinical investigations. We present an open source reference system for portable low-field MRI designed to support replication, reproducibility, benchmarking, and quantitative comparisons.
     
    The proposed MRI system is fully open source, based on a \qty{\sim50}{\milli\tesla} permanent-magnet and integrated with a cloud-native acquisition platform. Pulseq based system calibration, characterization and imaging sequences were used to evaluate the system's noise level, eddy currents, image-based SNR, and geometric accuracy. Quantitative \T1, \T2, and \B0 mapping sequences were developed and assessed using reference values. First results from independent system replications by two different sites were evaluated.
     
    The system achieved a noise level of \num{1.4} relative to the thermal noise floor, whereas the eddy current characterization yielded short decay constants of \qtyrange{27}{32}{\micro\second} across all gradient channels. Geometric deviations were $\leq$ \qty{2}{\milli\metre} over the investigated field of view. Noise levels and image-based SNR were comparable across the independent system replications. Measured \T1 values closely matched the specified values with an average absolute error of \qty{3.1\pm1.8}{\percent}, while \T2 values were overestimated by \qty{10.4\pm5.8}{\percent}. Simulations only showed marginal errors for both quantitative values, suggesting experimental error sources for \T2 mapping.
     
    The reference system combines openly documented hardware, software, calibration procedures, quality-control phantoms, quantitative MRI, and simulation tools within a reproducible ecosystem. By providing a fully characterized and openly accessible reference system, we aim to support cross-site comparability, reproducible research, and collaborative development of future portable low-field MRI technologies.
\end{abstract}

\keywords{
    Low-field \and MRI \and Halbach \and Portable \and Affordable \and Accessible \and Permanent magnets \and Open Source \and Reproducibility
}

\renewcommand{\thefootnote}{}\footnotetext{
    \textbf{Funding:} This work was supported by the project 22HLT02 A4IM which has received funding from the European Partnership on Metrology, co-financed from the European Union's Horizon Europe Research and Innovation Programme and by the Participating States. This research is funded by dtec.bw -- Digitalization and Technology Research Center of the Bundeswehr. dtec.bw is funded by the European Union -- NextGeneration EU.
}

\section{Introduction}\label{introduction}

MRI is an indispensable diagnostic tool known for its excellent soft tissue contrast and functional imaging capabilities. Nevertheless, conventional MRI is limited by substantial burdens, namely high acquisition and operating costs, infrastructure requirements for magnetic shielding and cooling, additional burden on the clinical workflows due to the fixed-installation, acoustic noise, and contraindications for individuals with implants or claustrophobia.

In response to these limitations, recent years have seen a surge in the development of low-field and portable MRI systems aimed at decentralizing access and enabling point-of-care imaging. Commercial systems have demonstrated feasibility for bedside brain imaging at \qty{64}{\milli\tesla} in intensive care settings, including assessment of intracranial midline shift\cite{sheth2022} and evaluation of intracranial hemorrhage\cite{mazurek2021}.
Developed within academic research projects, low-field MRI systems based on Halbach arrays, foster affordability and have much smaller footprints and are easier to assemble and reproduce. As they are build with permanent magnets they are cryogen-free and can be operated almost maintenance-free. In-vivo imaging results have been demonstrated with a \qty{50}{\milli\tesla} Halbach-system for brain and extremities\cite{oreilly2019, oreilly2021}, also yielding quantitative imaging results\cite{oreilly2021-2}.
A system which was optimized for physiological imaging at \qty{70}{\milli\tesla}, could be successfully deployed in various scenarios, such as patient homes, offices or outdoors.\cite{guallart-naval2022, algarin2023}
A modified Halbach array has been proposed\cite{cooley2021, cooley2015} which incorporates the frequency encoding gradient in the static magnet field, relaxing the hardware requirements by the gradient system. With use cases targeting neurological diseases, most of the systems are tailored to image the brain and extremities. Most recently, a whole body low-field MRI with \qty{50}{\milli\tesla} field strength has been demonstrated\cite{zhao2024}, extending the field of application to cardiac and abdominal imaging for instance.
Due to the much lower static magnetic field strength of point-of-care imaging devices, image quality is degraded by lower SNR and higher \B0 field inhomogeneities which causes geometric distortions and signal alterations. Besides, signal reception is more prone to electromagnetic interferences due to the lack of a Faraday cage.
However, several approaches to overcome these drawbacks have been implemented. Leveraging the availability of high performance computing hardware at lower costs, advanced deep learning-based reconstruction algorithms are able to improve SNR, while reducing geometric distortions. \cite{schote2024, lau2023} Additional information from auxiliary sensors\cite{srinivas2022, man2022} substantially reduce noise and artifacts from external electromagnetic interferences, while dedicated hardware modifications of the RF coil \cite{pfitzer2026, vliem2026} or the grounding and shielding strategy of the patient\cite{lena2026} help to initially suppress most external noise contributions.

This momentum has, however, not yet translated into a common technical open source baseline across different sites. Some open designs are so far limited to small bore systems\cite{block2025} not targeting clinical applications and published scanners are typically one-off research prototypes, where design files, bills of materials, assembly instructions and control software are released only in part, if at all, so that reproducing a scanner at a second site effectively amounts to redesigning it. The consequences are threefold. First, development effort is repeatedly spent on solved problems instead of new methods. Second, results obtained on individual builds remain difficult to compare, since neither the hardware configuration nor its characterization is shared in a form that allows independent verification and third, regulatory challenges for clinical studies need to be solved on an individual basis which amplifies the translational burden. Consequently, the open source documentation of an MRI system should include guidelines for typically safety testing, documentation for ethical approval or a full certification process which follows international standards and local regulations. Such documentation blueprints do not exist and are usually out of scope for research papers. 

This lack of a shared reference limits the ability to systematically evaluate design trade-offs between magnet topologies, RF chain architectures and image encoding strategies. Fundamental performance metrics\cite{guallart-naval2023}, such as \B0 inhomogeneity over a defined spherical volume, receive-chain noise and geometric accuracy, are often omitted or reported without a traceable measurement procedure and an associated uncertainty. Quantitative imaging offers a direct means to close this gap. Measurements of relaxation times and diffusion parameters on standardized phantoms with reference values can be compared both against these references and across sites. Associating a measurement uncertainty to those values allows for a rigorous assessment of the system reproducibility.

\begin{figure}[ht]
    \centering
    \includegraphics[width=0.72\textwidth]{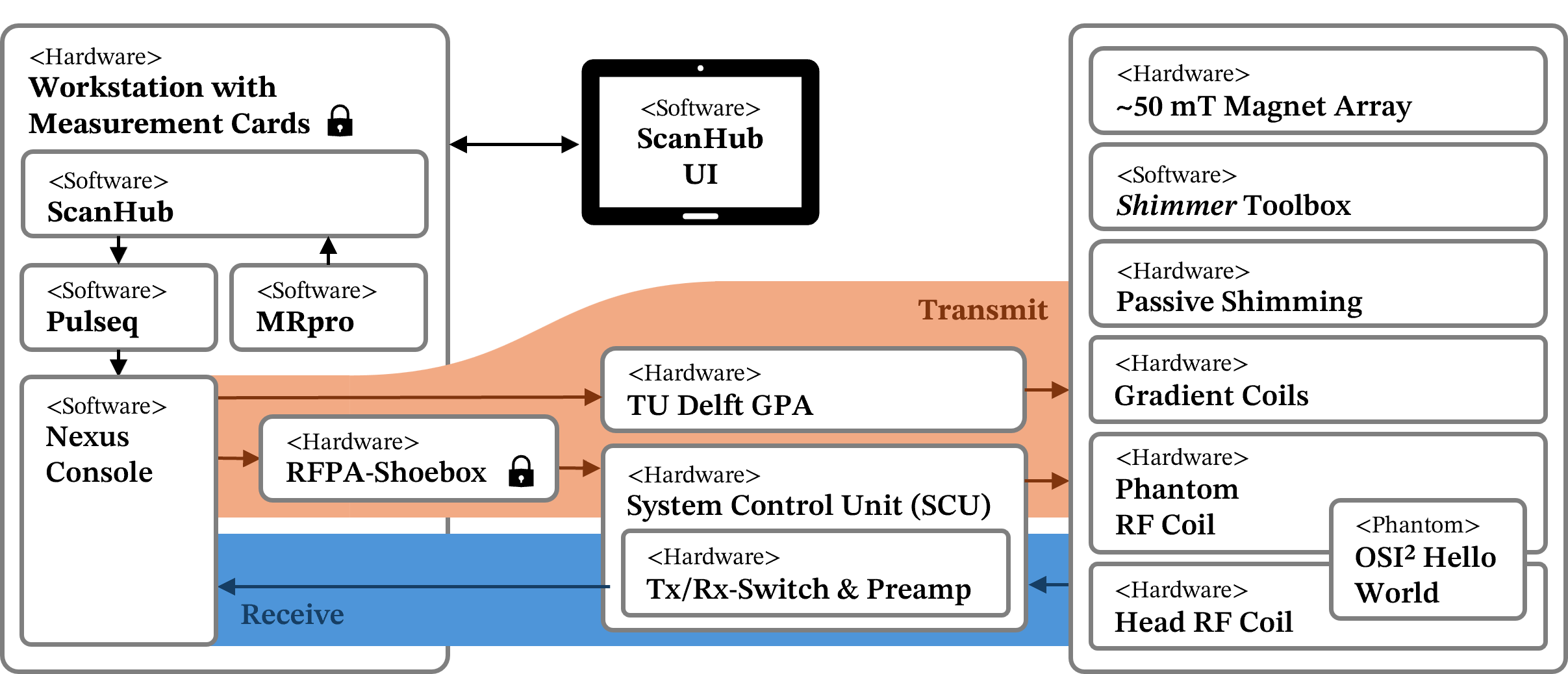}%
    \hfill%
    \includegraphics[width=0.26\textwidth]{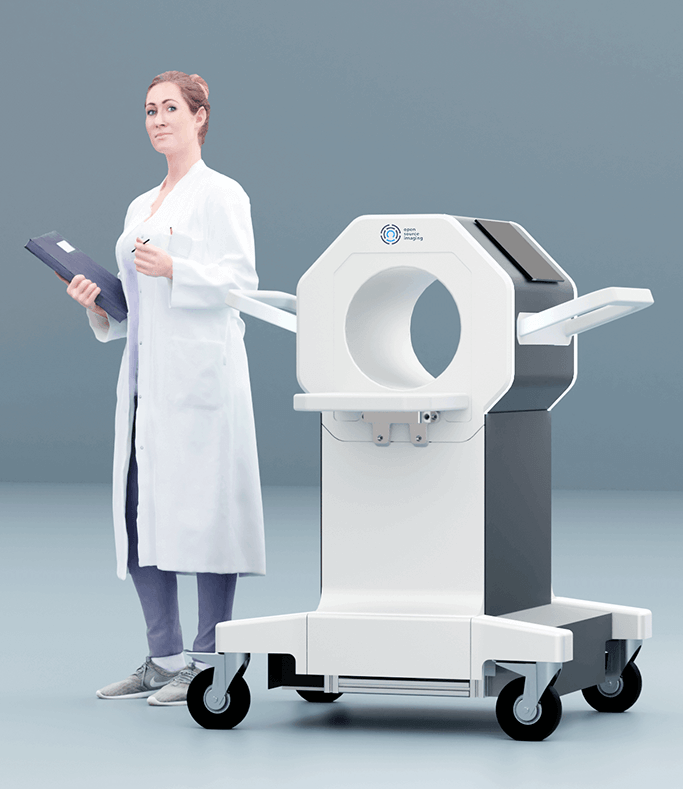}%
    \caption{Overview of the system built from open source components (left), apart from the workstation and RFPA, which contain closed-source hardware. The rendering (right) illustrates the result of a design study which demonstrated the compact integrability of the described system components in a medical MRI scanner.}\label{fig:system-architecture}
\end{figure}

The aim of this work is to establish a reproducible reference system whose hardware, software and system characterization is open source, such that the system can be rebuilt by others and independently verified. This provides a solid technological basis for scientific and industrial innovations that can ultimately be translated into medical products that improve healthcare.\cite{winter2024} An open source foundation can further reduce the financial and personnel resources required to construct, operate, maintain, and upgrade low-field MRI scanners,\cite{winter2019} thereby providing a global common basis for affordable medical technologies, which is urgently needed.\cite{anazodo2023} Here, we describe the individual hardware components of the system, their integration and the software which operates the system. In addition, we provide quantitative results which include the system characterization, multi-parametric imaging and MRI simulations for validation and cross-site comparisons. To complement the technological foundation of the scanner an extensive open source documentation is provided serving as a blueprint for safety testing according to IEC 60601-1\cite{iec60601-1_2020} and IEC 60601-2-33\cite{iec60601-2-33_2022} and for the certification as a medical device under the \gls{mdr} of the EU 2017/745\cite{eu_mdr_2017_745}. As such, the long-term vision is the ability to gather in-vivo imaging data and facilitate the translation into a clinical scanner to ultimately benefit patients as illustrated by the rendering in \figurename\,\ref{fig:system-architecture},

All the resources used for the MRI scanner described in this work, are available under \url{https://gitlab.com/osii}, which serves as the platform composed of reviewed repositories for reproducible and open source low-field MRI. In addition, the acquired data presented in this work and the sequence files are published with Jupyter notebooks to enable easy adaptability.\cite{herthum_2026_21807140}

\section{Methods}\label{methods}

\subsection{System Components}

\subsubsection{Magnet}\label{sec:magnet}

Unlike in conventional high-field systems, the static magnetic field was not generated using superconducting technology but with NdFeB (grade N52) permanent magnets. Instead of aligning them in a strict Halbach configuration, as shown in \cite{oreilly2019, oreilly2021}, the orientations of the individual permanent magnets were optimized in a so called rotation-optimized magnet (ROMA) design to further increase magnetic field homogeneity.\cite{roma_magnet,zanovelloVerylowfieldMRIScanners2025}.

We arranged 1936 cubic \qtyproduct{12 x 12 x 12}{\milli\metre} and 384 elongated \qtyproduct{50 x 12 x 12}{\milli\metre} magnets in a cylindrical geometry with a vertically oriented \B0 pointing. The final assembly had an inner diameter of \qty{322}{\milli\metre}, outer dimensions of \qtyproduct{450 x 520 x 567}{\milli\metre}, and a total mass of approximately \qty{100}{\kilo\gram}. The design\cite{osi2one-gitlab} focused on reproducibility, portability, and ease of assembly. The magnet holder rings were machined from polypropylene on three-axis CNC milling machines. The rings were positioned on brass rods and closed with lids without the need of any adhesives, what allowed a straightforward assembly of the magnet by one person. The construction is fully reversible, allowing the magnets to be removed, replaced, and reused as required.

Magnetostatic field simulations were performed to optimize magnet placement and field homogeneity over a \qty{200}{\milli\metre} \gls{dsv}, which achieved an initial \B0 inhomogeneity of \qty{408}{\ppm}. Due to imperfections in the production process of the rings, changes in magnet orientation and position due to magnetostatic forces\cite{zanovelloVerylowfieldMRIScanners2025}, size and remanence tolerances of the permanent magnets, the initially constructed magnet was an order of magnitude less homogeneous than expected by simulations achieving a \B0 inhomogeneity of \qty{34143}{\ppm} in a \qty{200}{\milli\metre} \gls{dsv}. To improve \B0 field homogeneity, we incorporated passive shim trays optimized through the open source toolbox called Shimmer\cite{kulikovShimmerPassiveOpenSource2026}. It allowed to iteratively optimize the positions and orientations of the permanent magnets to minimize the \B0 field inhomogeneity based on spherical magnetic field measurements. Field mapping for the shimming procedure was conducted using the 3D positioning system \textit{COSI Measure}\cite{hanOpenSource3D2017} in combination with a triaxial Hall probe (Model 460, LakeShore Cryotronics, Westerville) and a NMR field probe (Model PT2026 Precision Teslameter, Metrolab Technology SA.). After three iterations of \B0 shimming, the inhomogeneity was reduced to \qty{1602}{\ppm} in \qty{200}{\milli\metre} \gls{dsv}. At a room temperature of \qty{24.7}{\degreeCelsius}, an average \B0 of \qty{48.47}{\milli\tesla} was measured, corresponding to a Larmor frequency of \qty{2.06}{\mega\hertz}. 

\subsubsection{Gradient Subsystem}\label{sec:gradient-system}

The gradient coil designs were optimized using \emph{CoilGen}\cite{amreinCoilGenOpensourceMR2022}, an open source numerical tool for gradient coil optimization, targeting symmetric performance across all channels with a maximum inductance below \qty{300}{\micro\henry} to be compatible to the \gls{gpa}. To achieve low inductance and high sensitivities on all three channels, the orthogonal basis of the gradient system was rotated, such that the resulting axis are pointing along (X+Y), (X-Y), and Z axes.\cite{gradient_coils,littinOpenSourceGradient2026} The calculated theoretical gradient efficiencies were \qtylist[list-units=single]{0.39;0.40;0.39}{\milli\tesla\per\metre\per\ampere} across the channels. The resulting fully interconnected wire paths were exported as \emph{.stl} file from \emph{CoilGen}. Solid cylindrical coil bodies were obtained after processing the wire paths in the openly available software \textit{Blender}\cite{blender_website}. Three separate coil formers were 3D printed using a Prusa XL and PETG filament (Prusa Research a.s., Czech Republik). Each wire path was manually wound using individually insulated Litz-wire (Pack Litz Wire, Germany).

We used a three channel \gls{gpa} with \qty{\pm15}{\ampere} peak current and \qty{30}{\percent} duty cycle per channel, optimized for inductive loads ranging from \qtyrange[range-units=single]{15}{300}{\micro\henry}.\cite{degansTUDelftGradient2022} Bode plots and step response measurements confirmed stable amplifier behavior within the operational load range, with a bandwidth of \qty{7}{\kilo\hertz} and a current rise time of \qty{50}{\micro\second}. The \gls{gpa} was supplied by two power supplies providing \qty{\pm15}{\volt} (SM 18-50, Delta Elektronika, Zierikzee, The Netherlands).

\subsubsection{RF Subsystem}

Excitation and signal reception were performed with single-channel \gls{txrx} solenoid coil. The transmit RF waveform generated by the console was amplified by an \gls{rfpa} (RFPA-Shoebox, barthel HF-Technik GmbH, Aachen, Germany). The received signal amplification is described in Section \ref{sec:scu}. The housings of both RF coils were 3D-printed using PLA with \qty{0.1}{\milli\metre} layer height and \qty{20}{\percent} infill. Both coils featured feet that allowed positioning into the OSI\textsuperscript{2}~ONE rail system together with a 3D printed phantom holder, making coil and sample positioning repeatable across experiments.

The RF coil dedicated to the \textit{OSI\textsuperscript{2}~ONE \emph{Hello World}} phantom (described in section \ref{methods-qc}), was a circular solenoid with an inner diameter of \qty{150}{\milli\metre} (outer diameter \qty{155}{\milli\metre}), a winding length of \qty{145}{\milli\metre} and a housing length of \qty{190}{\milli\metre}.\cite{phantom_rf_coil} Twenty equidistant turns of Litz wire (Rupalit\textregistered~V155, Rudolf Pack GmbH \& Co. KG, Gummersbach, Deutschland) were laid into a helical groove of the printed cylinder. The resonance condition was obtained by a combination of variable and fixed tuning and matching capacitors. Tuning and matching was performed with the \emph{Hello World} phantom loaded and the coil positioned at the scanner isocenter, which yielded a reflection coefficient of approximately \qty{-35}{\dB} at the Larmor frequency.

For head-sized loads, we used a dedicated solenoid coil with an elliptical cross-section.\cite{head_rf_coil, zanovello2026} The coil's inner diameter measured \qtyproduct{233 x 213}{\milli\metre}, with a housing length of \qty{190}{\milli\metre}. Its 14 turns followed an elliptical path of \qtyproduct{242 x 222}{\milli\metre} over a winding length of \qty{125}{\milli\metre}. The turns were distributed non-uniformly along the coil axis, the pitch decreased from \qty{14.1}{\milli\metre} at the center to \qty{4.7}{\milli\metre} at the ends and was optimized to maximize the \B1 magnetic field homogeneity and RF coil sensitivity. The winding was divided into two segments, connected through parallel capacitors, to reduce conductor voltages and thus the sensitivity of the resonance to dielectric loading. We positioned the lumped elements on two printed circuit boards contacting the winding through pin connectors. The balanced matching network and the coaxial connector were located on the feed board. Segmenting capacitors and an additional parallel capacitor to fine tune the resonance frequency were located on the center board.

\subsubsection{System Control}\label{sec:scu}

For sequence execution and data acquisition we used the Nexus console\cite{schote2025,nexus_github} and integrated the measurement cards into a modular high-performance workstation. It not only served as MRI console, but also performed image reconstruction and hosted the ScanHub acquisition platform with a user interface described in Section \ref{sec:scanhub}. 
All sequences were implemented in the open Pulseq format\cite{laytonPulseqRapidHardwareindependent2017, raviPyPulseqPythonPackage2019}, which can be played directly on the console. The raw data were exported from the console in \emph{ISMRMRD} format\cite{inatiISMRMRawData2017}, which allowed direct integration with the existing MRI reconstruction framework \emph{MRpro}\cite{zimmermannMRproOpenPyTorchbased2025}.
A console configuration file defined the hardware specific parameters including \gls{gpa} gain, gradient coil efficiencies, a factor converting the \B1 field to a console output, and a look up table for non-linearity compensation of the \gls{rfpa} gain. 

In addition to the console, we designed a \gls{scu}\cite{frintzOpenSourceModularSystem2026} based on a standardized 19-inch, 3-unit, 84-HP Eurocard subrack (IEC~60297-3-100) hosting \qtyproduct{100 x 160}{\milli\metre} plug-in modules interconnected by a shared DIN~41612 backplane to provide power distribution, and inter-module communication. This avoided custom machining, simplified assembly and reduced the overall system footprint by integrating all control subsystems into a single shielded enclosure. We designed a custom four-layer backplane with IEC~60603-2 connectors, to distribute \qty{+24}{\volt} PELV, system reset, SPI with \qty{4}{\bit} address bus, UART, I\textsuperscript{2}C and emergency-stop signals. Power was supplied by an off-the-shelf industrial unit (13100-105, Schroff, Straubenhardt, Germany). RF signals were deliberately excluded from the backplane and routed coaxially on the front panel.
To minimize electromagnetic interference and preserve mixed-signal integrity, the PCBs incorporated dedicated internal ground planes that provided short return-current paths, internal shielding, and low-impedance grounding to the conductive chassis through ENIG-plated mounting points.\cite{guallart2026electromagnetic}

A fully passive \gls{txrx}-switch was integrated, which was designed around a quarter-wavelength impedance transformer following antiparallel diodes. The \gls{txrx}-switch isolated the pre-amplifier during transmission and allowed the MR signal to pass from the coil to the pre-amplifier with minimal loss during reception. The \gls{txrx}-switch and a two-stage \gls{lna} (PHA-13LN, Mini-Circuits, NY, USA; \qty{46}{\decibel} gain, \qty{3}{\decibel} noise figure) were integrated into a single Eurocard submodule. We used impedance controlled transmission lines and added two PCB shielding cans with removable lids. Bench measurements using an HP 8648C signal generator, a Keysight DSOX1204G oscilloscope, and the RFPA-Shoebox confirmed reliable pre-amplifier protection for the nominal \gls{rfpa} peak output power of \qty{100}{\watt} at a maximum duty cycle of \qty{10}{\percent}.

\begin{figure}[ht]
    \centering
    \includegraphics[width=0.5\linewidth]{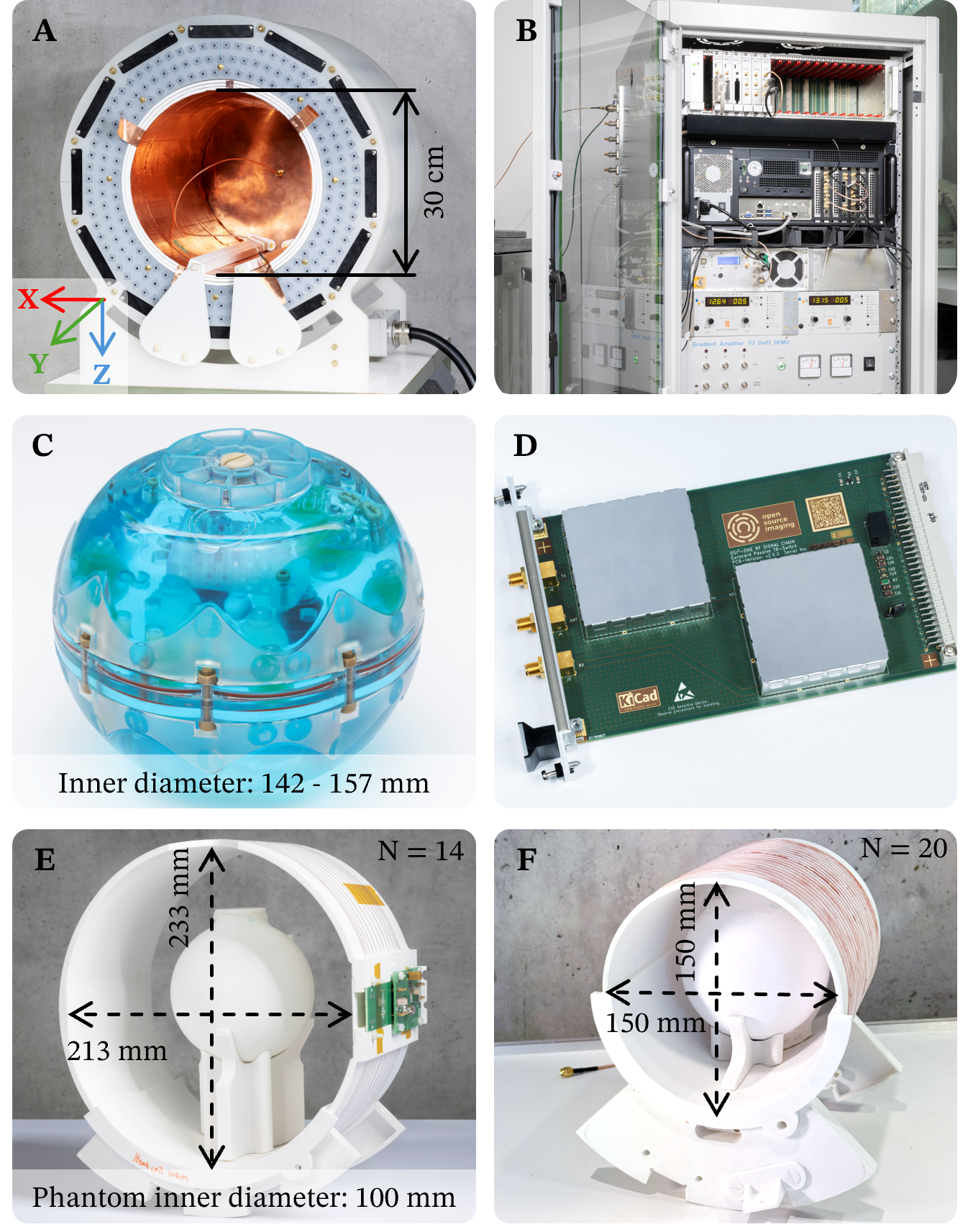}%
\caption{Photographs of the system components, including the magnet array in ROMA configuration (A), a rack integrating the SCU, console, RFPA and GPA (B), the \emph{CaliberMRI} phantom with quantitative reference values for \T1-, \T2- and diffusion mapping (C), the \gls{txrx}-switch module with an embedded pre-amplifier (D) an RF coil optimized for head imaging loaded with the \emph{Hello World} phantom (described in section \ref{methods-qc}) (E) and a solenoid coil for phantom imaging (F).}\label{fig:system-photographs}
\end{figure}

\subsubsection{Integration of Subsystems}\label{sec:integration}

The gradient coil set was first mounted inside the magnet array. Its terminals were routed to a connection box fitted with an industrial high-current plug (Han 16B series, HARTING Deutschland GmbH \& Co. KG, Minden, Germany). Using a single cable (ÖLFLEX® CLASSIC 110 CY BLACK 0,6/1 kV, U.I. Lapp GmbH, Stuttgart, Germany) and plug for all gradient lines avoided channel misassignment. 3D-printed adapters attached to the rear of the magnet prevented translation along the bore and rotation of the gradient coils, fixing the assembly isocentrically. A one-sided open Faraday cage was inserted as an RF shield inside the gradient coil, formed from a semi-rigid substrate with two layers of \qty{\approx7}{\micro\meter} copper, soldered into a cylinder. The shield was wired to a copper back-plate that carries the connection port for the RF coil and ties coil and shield to a common RF ground. Within the shield. The RF coil was mounted on a 3D-printed rail system to ensure reproducible positioning.\cite{rail_system} 

All electronic components, including the \gls{scu}, Nexus console, \gls{rfpa}, \gls{gpa} and power supply, were housed in a mobile 19-inch rack. A connection panel on one side wall interfaced the RF coil and gradients, while the opposite wall provided the \qty{230}{\volt}/\qty{16}{\ampere} mains inlet and a network port. Consolidating the electronics in a single rack enabled a well-defined common ground, clean cable routing and consistent shielding across all components. The overall build is affordable, weights less than \qty{200}{\kilo\gram}, is portable and can be operated using a standard power outlet.

\subsection{ScanHub}\label{sec:scanhub}
\subsubsection{User Interface}\label{sec:scanhub-ui}

\begin{figure}[ht]
    \centering
    \includegraphics[width=\linewidth]{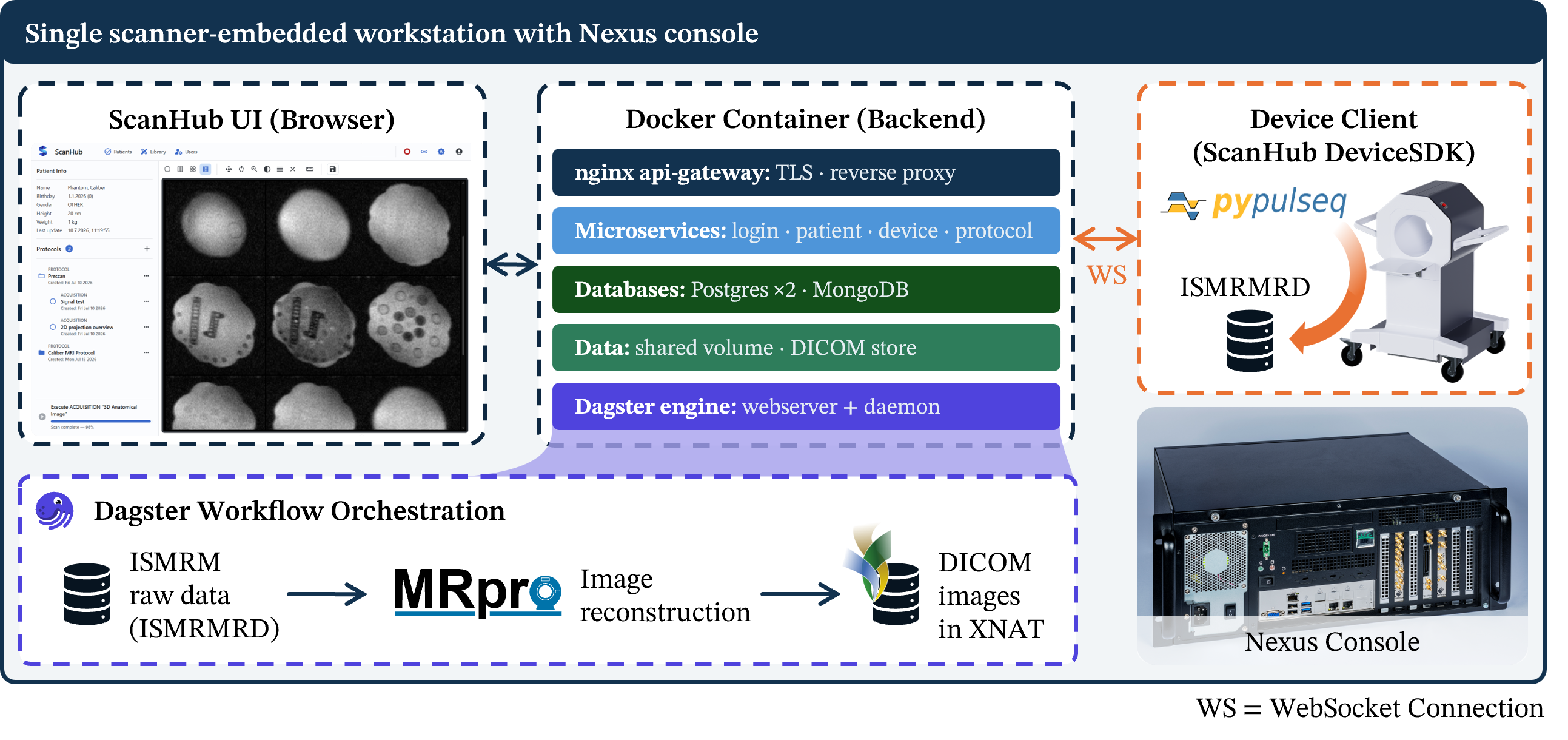}%
    \caption{ScanHub software embedded in the Nexus console workstation of the proposed low-field MRI system. The scanner is operated by the browser-based user interface, which is deployed via Docker Compose and runs as a container on the workstation. It allows to visualize \emph{DICOM} images and trigger acquisitions. The microservice-based backend for patients, protocols, sequences, devices and workflow orchestration provides end-to-end data governance, which can also be shared across devices and can be distributed to different compute nodes if the workload increases.}\label{fig:scanhub-integration}
\end{figure}

To make the scanner operable by a user, we adopted and further developed \emph{ScanHub}, an open source acquisition platform that embeds the scanner into a standardized, cloud-native ecosystem. \cite{scanhub_github,schoteRemoteControlPortable2026,schoteScanHubOpenSourcePlatform2023} We employed a microservice-based architecture (Figure~\ref{fig:scanhub-integration}), with services distributed across \emph{Docker} containers that shared a common base image to enforce type-safe and consistent interfaces. Integrated services included user-login management, patient management, protocol management, device management, and a workflow orchestration engine. The microservices were implemented in \emph{Python} using the \emph{FastAPI} framework, with their HTTPS/REST endpoints exposed through an \emph{Nginx} reverse proxy. Structured data were kept in two \emph{PostgreSQL} instances, pulse sequences in \emph{MongoDB}, and raw data, images, and pipeline artifacts in a shared data lake. The user interface was implemented with \emph{React}, \emph{TypeScript}, and \emph{Vite} such that it runs directly in a web browser.

For the examinations we implemented a tree structure of patients or subjects, protocols, and acquisition tasks, instantiated from reusable templates to ensure protocol reproducibility. Each acquisition task referenced a \emph{Pulseq} sequence, the target device, and acquisition parameters, e.g. the field of view, and calibration routines. Acquired data were reviewed in integrated viewers for reconstructed \emph{Dicom} images and \emph{ISMRMRD} files. A device SDK enabled arbitrary console hardware to be connected to \emph{ScanHub} via callback functions defining scans and calibrations. The stack was deployable with a single \emph{Docker Compose} command on the \emph{Nexus} console workstation. The system was accessible for users from any browser on the network, including portable devices such as tablets.

\subsubsection{Image Reconstruction Workflow}\label{sec:scanhub-reco}

Within \emph{ScanHub}, image reconstruction was delegated to an embedded workflow orchestration engine based on \emph{Dagster}, in which each processing step was defined as a data asset and pipelines were expressed as directed acyclic graphs. The completion of the raw data transfer from the console automatically triggered the submission of the reconstruction job through the \emph{Dagster} API. The run configuration carried the task identifier, the task directory and the access token, so that every run was linked to the corresponding acquisition task. The reconstruction job consisted of two assets. The first loaded the raw data and the second performed the image reconstruction with \emph{MRpro}\cite{zimmermannMRproOpenPyTorchbased2025} using the Cartesian trajectory calculator and the direct reconstruction method. A dedicated I/O manager stored the resulting image data as \emph{DICOM} files. Run-status sensors closed the loop to the platform and propagated the status to the user interface. 

\subsection{System Calibration}\label{methods-calibration}

Prior to each measurement, a calibration routine was used to adjust the Larmor frequency, the RF transmit power and the first-order shim currents to the gradient coils. After the RF coil has been manually tuned to the expected Larmor frequency, the Larmor frequency was estimated from the decaying half of a spin-echo signal, acquired with a bandwidth of \qty{20}{\kilo\hertz} and an \gls{adc} duration of \qty{50}{\milli\second} to yield a spectral resolution of \qty{20}{\hertz} per point. The transmit frequency was then set to the peak of the magnitude spectrum. Subsequently, the strength of the \B1 pulse was calibrated by acquiring repeated \gls{fid} signals with a long TR of \qty{3}{\second} and stepwise increasing transmit power. The long TR ensured that magnetization recovered to equilibrium between excitations even when samples with long \T1 relaxation times are used. The transmit scaling factor was selected such that a nominal flip angle of \ang{90} produces the maximum signal. As the required power depends on the coil load, this step was repeated whenever the phantom or subject was exchanged. Using the calibrated transmit power, the gradient offset currents were adjusted for first-order active \B0 shimming. Following 3 dummy scans, repeated \gls{fid} signals were acquired in magnetization steady-state with a TR of \qty{830}{\milli\second} and a reduced flip angle of \ang{45} to avoid overtipping. During the acquisition the shim offsets were swept over a predefined range of \qtyrange{0.01}{0.1}{\milli\tesla\per\metre\per\ampere}. For each step size, the positive and negative offset is examined successively for each channel. The combination which yielded the maximum \gls{fid} amplitude was retained.

\subsection{System Characterization}\label{methods-performance}

\subsubsection{Noise}\label{methods-noise}

To characterize the noise of the receive chain and to isolate the contribution of \gls{txrx}-switch (including the preamplifier) and the coil, noise data (i.e., without RF excitation) was acquired in a series of system configurations.\cite{guallart2026electromagnetic} To ensure, that the noise level is not limited by the \gls{adc}, the \gls{adc} input of the console was terminated with \qty{50}{\ohm}. Subsequently, the \gls{txrx}-switch together with the low-noise preamplifier was added, terminating the coil port with \qty{50}{\ohm}, and finally connected the RF coil tuned to the Larmor frequency and loaded with the \emph{Hello World} phantom. In this fully assembled configuration, noise was additionally acquired during an active readout gradient applied separately on each of the three gradient axes to assess the noise contribution of the gradient chain.
For each configuration, three repetitions were acquired with a TR of \qty{500}{\milli\second}, an \gls{adc} duration of \qty{100}{\milli\second} and an \gls{adc} bandwidth of \qty{20}{\kilo\hertz}. To compare noise with and without an activate gradient, a trapezoidal readout gradient with a maximum amplitude of \qty{4.7}{\milli\tesla\per\metre} was used.

The noise of each configuration was quantified by the RMS of the real part of the acquired signal. This value is normalized by the theoretical Johnson--Nyquist thermal noise floor to obtain a dimensionless noise level,
\begin{equation}\label{eq:noise-level}
    \mathrm{Noise\,level} = \frac{\mathrm{RMS}}{\nu_{\mathrm{out}}}.
\end{equation}
A value close to unity corresponds to a receive chain that is limited by the thermal noise of the source resistance. As derived in Guallart-Naval et al.\cite{guallart2026electromagnetic} the thermal noise voltage $\nu_{\mathrm{out}}$ is computed as
\begin{equation}\label{eq:johnson}
    \nu_{\mathrm{out}} = G \, \mathrm{NF} \, \sqrt{k_{\mathrm{B}} \, T \, R \, \Delta f}.
\end{equation}
with the Boltzmann constant $k_{\mathrm{B}} = \qty{1.38e-23}{\joule\per\kelvin}$, room temperature $T = \qty{300}{\kelvin}$, coil impedance matched to $R = \qty{50}{\ohm}$ and \gls{adc} bandwidth $\Delta f = \qty{20}{\kilo\hertz}$. We considered the total gain $G = 199.5\,(\qty{46}{\decibel})$ and noise factor $\mathrm{NF} = 1.4\,(\qty{3}{\decibel})$ of our two-stage pre-amplifier design. Note that equation \eqref{eq:johnson} already accounts for the halving of the open-circuit thermal noise voltage at the matched \qty{50}{\ohm} preamplifier, yielding $\nu_{\mathrm{out}} = \qty{18.13}{\micro\volt}$.

\subsubsection{Eddy Currents}\label{sec:ec-methods}

The RF shield placed between the gradient and RF coils introduced in Section~\ref{sec:integration} is supposed to suppresses external noise coupling into the receive chain, but its conductive layers are subject to gradient-switching-induced eddy currents that can cause image artifacts.\cite{devosSegmentedRFShield2024} To minimize this effect,  a thin, semi-rigid two-layer laminate with \qty{18}{\micro\metre} of copper on each side was used. By deriving the time constants we aim to characterize the systems eddy currents arising from gradient switching with a small pickup coil (radius $r = \qty{35}{\milli\metre}$, $N = 3$ turns) to assess the impact on the pulse sequence design. According to Faraday's law of induction, the time-varying flux $\Phi = B\,A$ in the coil induces the voltage
\begin{equation}
    V_{\mathrm{ind}} = -N\,\frac{\mathrm{d}\Phi}{\mathrm{d}t} = -N\,A\,\frac{\mathrm{d}B}{\mathrm{d}t},
\end{equation}
with coil area $A = \pi r^{2}$. Since the gradient field varies linearly with position, $B(x) = G\,x$, the field rate at the coil equals the slew rate scaled by the distance from the isocenter, $\mathrm{d}B/\mathrm{d}t = x\,\mathrm{d}G/\mathrm{d}t$. For the measurements, a trapezoidal gradient waveform representative of the gradients used in an imaging sequences was applied, with a slew rate of \qty{50}{\tesla\per\metre\per\second} and a maximum amplitude of \qty{11}{\milli\tesla\per\metre}. We performed the measurement on the positive and negative end of each gradient within the bore, \qtyrange{100}{120}{\milli\metre} from isocenter, where the gradient field corresponds to \qtyrange{1.1}{1.32}{\milli\tesla} what translates to an expected peak induced voltage of \qtyrange{-57.7}{-69.3}{\milli\volt}.

Without eddy currents, $V_{\mathrm{ind}}$ would reproduce the rectangular derivative of the rising or falling gradient slope and vanish on the plateau. The superimposed, decaying eddy current field instead adds an exponential tail after each ramp. The pickup-coil response to the gradient drive current $I(s)$ was modeled as a first-order, linear time-invariant system,
\begin{equation}\label{eq:ec-lti-system}
    V_{\mathrm{ind}}(s) \approx H(s)\,I(s), \qquad H(s) = K\,\frac{s}{1 + \tau s},
\end{equation}
where the numerator $s$ represents the pick-up coil's differentiating ($\mathrm{d}B/\mathrm{d}t$) response, while the first-order pole approximates the dominant eddy-current time constant $\tau$, and $K$ scales the amplitude. In time domain, the step response of this system is a mono-exponential decay of the induced voltage $ V_{\mathrm{ind}}(t) \propto e^{-t/\tau}$. This response was fitted using \texttt{curve\_fit} from \emph{SciPy}\cite{virtanen2020scipy} on the falling ramp of each gradient axis and compare the corresponding time constants $\tau$.

For the measurement, the pickup coil was positioned manually, so that its exact location and orientation relative to the gradient coil are not quantified. This affects only the coupling between the gradient field and the coil, which enters \eqref{eq:ec-lti-system} as the gain factor $K$, and leaves the time constant $\tau$ unaffected. For the evaluation, we report the \num{99}th percentile instead of the peak value to obtain a robust amplitude measure.

\subsubsection{SNR and Geometric Accuracy}\label{methods-qc}

To benchmark system performance and support component calibration, a reproducible \emph{Hello World} phantom was developed. The phantom consists of a 3D-printed cylindrical shell (\qty{100}{\milli\metre} inner diameter, \qty{110}{\milli\metre} outer-diameter) filled with deionized water and $\mathrm{CuSO_4}$ (\qty{1.5}{\gram\per\liter}). Internal asymmetries and printed orientation-letters L (patient left), R (patient right), S (supine), I (inferior), A (anterior), P (posterior) facilitate the determination of unambiguous orientation. The physical gradient directions together with the selected slice orientation (coronal, axial, sagittal) define the \emph{ISMRMRD} header information such that the exported \emph{DICOM} files are displayed correctly. Moreover, embedded structures provided contrast and homogeneous regions were used for the evaluation of SNR and geometric accuracy.\cite{hello_world}

For consistent placement, a modular rail and mounting system was designed,\cite{rail_system} aligning the phantom within the RF coil and ensuring repeatable positioning within the magnet’s \gls{dsv}, see Figure~\ref{fig:system-photographs}. Imaging was conducted using a 3D \gls{tse} sequence with voxel size = \qtyproduct{2 x 2 x 2}{\milli\metre} and readout direction from left to right. Further imaging parameters were \gls{fov} = \qtyproduct{140 x 140 x 140}{\milli\metre}, TR = \qty{600}{\milli\second}, TE = \qty{12}{\milli\second}, \gls{etl} = 7, readout \gls{bw} = \qty{20}{\kilo\hertz}. Each scan was preceded by a \qty{1}{\second} noise acquisition (only \gls{adc} gate open) with the same readout bandwidth. 

The SNR was determined from the ratio of the mean phantom signal and the RMS value of the real part of the complex noise samples. In addition, a noise profile was derived from the noise scan to assess the RF receive profile. Although noise-profile normalization can compensate for the RF receive profile, it was not applied here due to the low readout bandwidth used in the experiments. To determine the geometric accuracy of the images, the S-I and L-R direction were manually determined by measuring the phantom size for a central sagittal and axial slice. The accuracy along the A-P direction was not determined due to the air bubble which formed at the anterior location of the phantom. 

\subsection{Quantitative MRI}\label{methods-qmri}

Quantitative imaging of \T1 and \T2 relaxation times and \B0 mapping was performed in addition to structural imaging with \gls{pdw} and \gls{t1w} using a commercial phantom tailored to \qty{50}{\milli\tesla} field strength (CaliberMRI, Inc. (formerly HPD/QalibreMD), in Boulder, CO USA, Low Field Single Plate Phantom Model 139, serial number: 139-0003).\cite{martin2023relaxation} The phantom had an outer diameter of \qty{170}{\milli\metre} while the fluid filled part had a maximum diameter of \qty{157}{\milli\metre} and a minimum diameter of \qty{142}{\milli\metre}.
The phantom contained 14 spherical inclusions (diameter \qty{17}{\milli\metre} surrounded by \qty{1.5}{\milli\metre} thick encapsulation) in a central section (center line + \qty{4}{\milli\metre}) of the phantom. Five inclusions have specified values of \T1 and \T2 times. Each inclusion was masked manually using the reconstructed parameter maps to calculate the mean and standard deviation of the respective relaxation times within the inclusion. The phantom is shown in Figure~\ref{fig:fig_anat_caliber}.

Images were acquired with an isotropic \gls{fov} of \qtyproduct{200 x 200 x 200}{\milli\metre}, readout oversampling factor 2, readout presampling with 10 samples and readout direction from left to right. For \gls{pdw} imaging a 3D \gls{tse} sequences with the following parameter was used: voxel size = \qtyproduct{2 x 2 x 2}{\milli\metre}, TR = \qty{600}{\milli\second}, TE = \qty{14}{\milli\second}, \gls{etl} = 10, readout \gls{bw} = \qty{20}{\kilo\hertz}, 4 dummies excitations, duration = \qty{473}{\second} and undersampling of the corners of k-space. For \gls{t1w} imaging, the TR was reduced to \qty{210}{\milli\second} with \gls{etl} = 5 and duration = \qty{330}{\second}. The noise acquisitions from \ref{methods-noise} and the mean phantom signal with \gls{pdw} and \gls{t1w} were used to calculate the SNR for the respective structural scans. Moreover, a digital twin of the phantom was created to perform MRI simulations to support sequence development and quantitative image reconstruction as outlined further below and documented in the respective repository.\cite{digital_twin} 

\subsubsection{\T1 and \T2 Mapping}

For T1 mapping, the 3D \gls{tse} sequence was preceded with an adiabatic inversion pulse\cite{kupce1995stretched} and varying TI = \qtylist[list-units=single]{50;100;150;230;350;600}{\milli\second}. Further imaging parameters were voxel size = \qtyproduct{3.4 x 3.4 x 12.5}{\milli\metre}, TR = \qty{2000}{\milli\second}, TE = \qty{10}{\milli\second}, \gls{etl} = 8, readout \gls{bw} = \qty{20}{\kilo\hertz}, duration = \qty{1392}{\second}.  For \T2 mapping, a 3D multi-echo spin-echo sequence was used to acquire a series of multi-echo images corresponding to the \gls{etl}. Further imaging parameters were: Voxel size = \qtyproduct{4.5 x 4.5 x 20}{\milli\metre}, TR = \qty{1000}{\milli\second}, TE from \qty{11}{\milli\second} to \qty{154}{\milli\second} in steps of \qty{11}{\milli\second} (\gls{etl} = 14), readout \gls{bw} = \qty{20}{\kilo\hertz}, duration = \qty{440}{\second}. The individual images at different inversion times and echo times were reconstructed using \emph{MRpro} with subsequent model-based reconstruction of \T1 and \T2 relaxation times using dictionary matching implemented in \emph{MRpro}.

\subsubsection{\texorpdfstring{\B0}{B0} Mapping}
 
\B0 mapping was performed for the \emph{CaliberMRI} phantom, and additionally for a spherical phantom with a diameter of \qty{165}{\milli\meter} filled with deionized water and $\mathrm{NiSO_4 x 6H_20}$ (\qty{1.25}{\gram\per\liter}), resulting in relaxation times of \T1 = \qty{\sim315}{\milli\second} and \T2 = \qty{\sim300}{\milli\second}. For both phantoms, the same 3D \gls{tse} sequence as for structural imaging was repeatedly acquired but with an increasing time shift between the peak spin echo formation and the beginning of the readout gradient, such that spin dephasing due to \B0 offsets could be reconstructed using two or more time shifts according to 
\begin{equation} 
\phi(\mathbf{r}, t) = \phi_0(\mathbf{r}) + \gamma \,\Delta B_0(\mathbf{r}) \cdot t.
\end{equation}
Here, $r$ is the coordinate locations, $t$ is time, $\gamma$ is the gyromagnetic ratio, $\Delta B_0$ is the \B0 field offset and $\phi_0$ is an initial constant phase offset. Images were acquired with time shifts of \qtylist[list-units=single]{0;0.75;1.5}{\milli\second}. Further imaging parameters were voxel size = \qtyproduct{5 x 5 x 5}{\milli\metre}, TR = \qty{1000}{\milli\second}, TE = \qty{10}{\milli\second}, \gls{etl} = 16, readout \gls{bw} = \qty{40}{\kilo\hertz}, 4 dummies excitations, duration = \qty{249}{\second}.

Complex images at all echo shifts were reconstructed using \emph{MRpro}. The phase of the images was further used to derive the underlying \B0 map, by analyzing the voxel-wise phase evolution over time. Therefore the phase images were temporally unwrapped using a quality-guided region-growing unwrapping algorithm \cite{Herraez:02, Dymerska_2021} and the unwrapped phase evolution was approximated with a linear regression model according to the equation above. \B0 maps with and without first-order active shimming for both phantoms were acquired, see Figure~\ref{fig:fig_qmri} and \ref{fig:fig_b0}. The added shim fields were validated by subtracting both field maps and comparing the residual shim gradients to the shim gradients set by the console. 

Moreover, the \B0 maps with and without active shimming of the spherical phantom were used to compute the \B0 inhomogeneity in ppm for increasing \gls{dsv} ranging from 10 to 160 mm diameter. The results were compared to previously acquired field probe measurements which were recorded during the initial passive shimming of the magnet after assembly.

\subsubsection{MRI Simulations}\label{methods-simulations}

To support sequence development a digital twin\cite{digital_twin} of the \emph{CaliberMRI} phantom was created to perform  MRI simulations using \emph{MRzero}\cite{loktyushin2021mrzero}. The same 14 inclusions (diameter = \qty{17}{\milli\metre}) were distributed accordingly and each inclusion was surrounded by a \qty{1.5}{\milli\metre} thick encapsulation with zero proton density, mimicking the plastic cover of the real inclusions. Each voxel was assigned a value for proton density, \T1 in \si{\second}, \T2 in \si{\second}, ${T_2}'$ in \si{\second}, diffusion coefficient D in \si{\micro\meter\squared\per\milli\second}, and \B0 inhomogeneity in \si{\hertz}, no \B1 inhomogeneity was added. Where available, the same values for \T1, \T2 and D as for the real phantom were used. The unknown values for \T1 and \T2 were assigned based on our previous measurements.\cite{herthum_qmri_ismrm2026} ${T_2}'$ was everywhere $0.7 \times \mathrm{T}_2$ and the measured \B0 map was used as \B0 inhomogeneity, see Figure~\ref{fig:fig_sim}.

For simulations, the same sequences as for \T1 and \T2 mapping were used. However, to reduce the computational load, the simulation was restricted to a single slice containing the inclusions. The simulated k-space was reconstructed using \emph{MRpro} with subsequent \T1 and \T2 mapping. Mean values and standard deviations were computed within the respective inclusions and compared to the measurements and vendor specified values.

\subsection{System Replication}\label{methods-reproducibility}

The previously described resources were already used to successfully replicate a portable low-field MRI system, amongst others, at TU Graz and \gls{inrim}. Admittedly, it was not possible to create exact copies of the reference system, partially, different open source components were chosen or self-designed modules were developed to accommodate the locally available resources. 

At TU Graz, the rebuild was successfully conducted over a one-year period based on the presented work. One goal of the rebuild involved using only publicly available resources. When possible, troubleshooting and questions were channeled through public forums, specifically, the \textit{OSI\textsuperscript{2}} Matrix Space (\url{https://matrix.to/#/#osii:matrix.org}) and \emph{GitLab} issues for respective modules\cite{osii_website}.

The rebuild included previously described modules like the main magnet (<\qty{3000}{\ppm} over a \qty{200}{\milli\metre} \gls{dsv}), the RF coils, the \gls{gpa} and the \emph{Hello World} phantom. Nevertheless, different open source designs for the console\cite{Negnevitsky_2023, osii_console}, the \gls{rfpa}\cite{osii_rfpa} as well as the active \gls{txrx} switch\cite{a4im_txrx} were used. 
Moreover, self-designed modules were used for the \gls{lna} (gain = \qty{45}{\dB}, noise figure ~ \qty{1}{\dB}) and for 3D printable gradient coils with gradient efficiencies of \qtylist[list-units=single]{0.44;0.91;0.60}{\milli\tesla\per\metre\per\ampere} along channels x, y, and z.

To evaluate the relative performance of the scanner variants, image-based SNR as well as geometric accuracy were evaluated for the TU Graz system using the \emph{Hello World} phantom (version without printed orientation letters). The experiments were carried out as previously described in section \ref{methods-qc}. However, the RF coil for the \emph{Hello World} phantom employed three capacitive segmentation points which were not included for the reference system. The RF shielding used a \qty{1}{\mm} thick copper shield between gradient coils and RF coil compared to the much thinner shield in the reference configuration. Additionally, the implementation for the 3D \gls{tse} sequence differed as no crusher gradients were employed and the data was fully sampled instead of undersampling the k-space corners. Due to differences in the gradient system, the TE was prolonged to \qty{15}{\ms} instead of \qty{12}{\ms}, as shorter echo times produced severe image artifacts. Other sequence parameters were kept identical (\gls{fov} = \qtyproduct{140 x 140 x 140}{\milli\metre}, TR = \qty{600}{\milli\second}, TE = \qty{15}{\milli\second}, \gls{etl} = 7, readout \gls{bw} = \qty{20}{\kilo\hertz}).

At \gls{inrim}, the replicated system configuration matched more closely the reference system as next to the main magnet, head-sized RF coil and \gls{gpa}, also the same gradient coils, \gls{txrx} switch, and \gls{lna} were used. However, the console and \gls{rfpa} were the same as at TU Graz. Besides, the \gls{inrim} system included a \SI{0.3}{\milli\meter} thick copper shield. The imaging  experiments at \gls{inrim} were carried out on the \emph{CaliberMRI} phantom using the same sequence \gls{pdw} TSE sequence as described previously but with a TE equal to \qty{14}{\milli\second},  without crusher gradients and no undersampling. The phantom was centered with respect to the coil through a 3D printed holder and the coil was positioned inside the magnet array by means of the rail system also used in the reference scanner.\cite{rail_system}

\section{Results}\label{results-performance}
\subsection{System Characterization}

\subsubsection{Noise}

The noise RMS and the noise level from \eqref{eq:noise-level} was evaluated per repetition and the mean and standard deviation over the three repetitions are reported in Table \ref{tab:noise-results}. The standard deviation of the noise level across repetitions remained below \num{0.06} in all configurations. Connecting the \gls{txrx}-switch and preamplifier and terminating the coil port with \qty{50}{\ohm} raises the noise level to \num{1.15} (A), as the thermal noise of the source resistance is amplified by the receive chain and combined with the noise generated by the \gls{lna}. Completing the chain with the tuned, phantom-loaded RF coil (B) increases the noise level by \qty{21.7}{\percent} to \num{1.40}, reflecting the additional contributions such as residual electromagnetic pickup. Applying a readout gradient during the \gls{adc} window (C, D, E) produced a small but systematic increase of the noise level across all three gradient axes. Relative to the end-to-end configuration without gradient (\num{1.40}), the noise level increased by \qty{7.1}{\percent}, \qty{0.7}{\percent} and \qty{9.3}{\percent} for axes~1, 2 and~3, reaching \num{1.50}, \num{1.41} and \num{1.53}, respectively. The increase is consistent in sign across all axes and exceeds the repetition-to-repetition scatter, indicating a small gradient-related noise contribution during active readout encoding. This contribution nevertheless remains minor compared with the dominant noise added by scenarios (B) and (C).


\begin{table}[ht]
    \centering%
    \caption{
        RMS and noise level calculated by \eqref{eq:noise-level} for each scenario. Values are given as the mean and standard deviation over the three repeated measurements.
    }\label{tab:noise-results}%
    \renewcommand{\arraystretch}{1.15}%
    \begin{tabularx}{\linewidth}{@{}l >{\raggedright\arraybackslash}X S[table-format=2.2(1.2)] S[table-format=1.4(1.2)]@{}}
        \toprule
        & Configuration & {RMS [\unit{\micro\volt}]} & {Noise level}\\
        \midrule
        (A) & TxRx-switch and preamp (\qty{50}{\ohm} at coil) & 20.80(0.13) & \num{1.15(0.01)} \\
        (B) & (A) + RF coil & 25.41(0.11) & \num{1.40(0.01)} \\
        (C) & (B) + readout gradient (Ch 1) & 27.17(0.88) & \num{1.50(0.05)} \\
        (D) & (B) + readout gradient (Ch 2) & 25.60(0.23) & \num{1.41(0.01)} \\
        (E) & (B) + readout gradient (Ch 3) & 27.70(0.53) & \num{1.53(0.03)} \\
        \bottomrule
    \end{tabularx}
\end{table}

\subsubsection{Eddy Currents}

\begin{table}[ht]
    \centering%
    \caption{
        Eddy current characterization of the three gradient axes for both gradient polarities. The time constant $\tau$ was obtained by measuring the voltage $V_{\mathrm{ind}}$ induced in a small pickup coil and fitting a first-order, linear time-invariant system defined in \eqref{eq:ec-lti-system}. $Q_{99}(|V_{\mathrm{ind}}|)$ denotes the \num{99}th percentile of the induced voltage magnitude, evaluated over the falling gradient ramp and $\varepsilon$ denotes the RMS fit residual normalized by $Q_{99}(|V_{\mathrm{ind}}|)$.
    }\label{tab:ec-results}%
    \begin{tabular*}{\linewidth}{@{\extracolsep{\fill}}l S[table-format=2.0] S[table-format=2.2] S[table-format=1.2]}%
        \toprule
        {Axis}
        & {$\tau \mathbin{/} \unit{\micro\second}$}
        & {$Q_{99}(|V_{\mathrm{ind}}|) \mathbin{/} \unit{\milli\volt}$}
        & {$\varepsilon \mathbin{/} \unit{\percent}$} \\
        \midrule
        $x+$ & 27 & 50.37 & 0.37 \\
        $x-$ & 30 & 48.19 & 0.96 \\
        $y+$ & 32 & 42.27 & 0.50 \\
        $y-$ & 29 & 38.31 & 0.53 \\
        $z+$ & 30 & 63.04 & 0.29 \\
        $z-$ & 29 & 59.13 & 0.39 \\
        \bottomrule
    \end{tabular*}%
\end{table}

\begin{figure}[ht]
    \centering
    \includegraphics[width=0.7\linewidth]{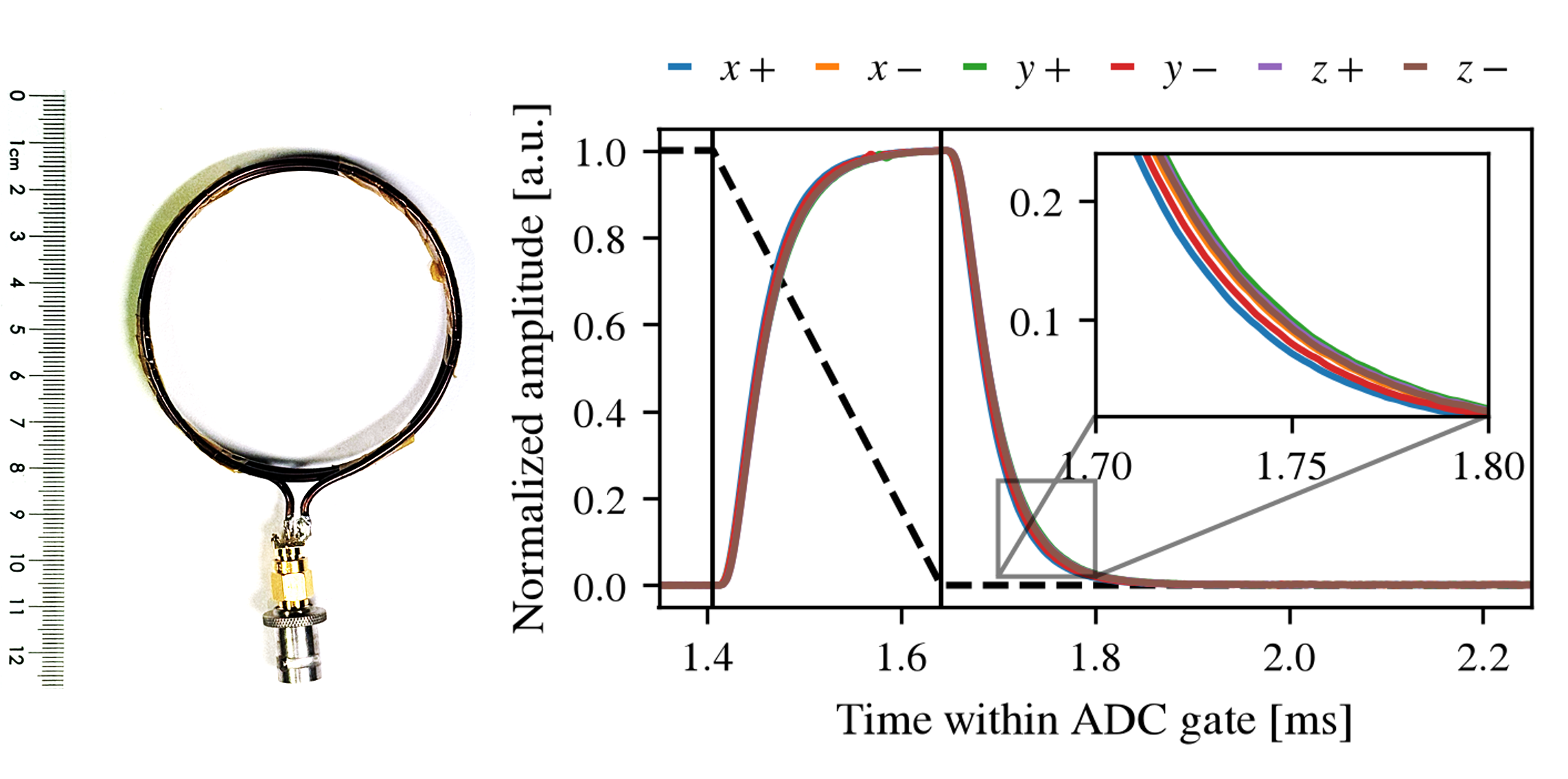}%
    \caption{Eddy current characterization from measurement with a pickup coil.}\label{fig:ec-results}
\end{figure}

Eddy currents were characterized for each gradient axis and both gradient polarities. Table~\ref{tab:ec-results} summarizes the time constants $\tau$ obtained from the fit of the first-order linear time-invariant system defined in \eqref{eq:ec-lti-system}, together with the \num{99}th percentile $V_{99}$ of the induced voltage magnitude evaluated over the falling gradient ramp. On this basis, the measured amplitudes agree with the values estimated in Section~\ref{sec:ec-methods}. The normalized RMS fit residual $\varepsilon$ in Table \ref{tab:ec-results} remains below \qty{1}{\percent} for all axes and polarities, indicating that the first-order model accurately represents the measured response.

Figure~\ref{fig:ec-results} shows the pickup coil used for the measurement together with the fits of the corresponding voltages measured at the pickup coil. The vertical lines delimit the falling gradient ramp used for the fit, the dashed line indicates the gradient current waveform, and the colored traces correspond to the six combinations of gradient axis and polarity. Note that Figure~\ref{fig:ec-results} plots the magnitude of the normalized traces to ensure comparability. All fitted time constants lie within \qtyrange{27}{32}{\micro\second}, and the difference between the two polarities of a given axis does not exceed \qty{3}{\micro\second}. The eddy current response is thus comparable across all three axes.


\subsubsection{SNR and Geometric Accuracy}

\begin{figure}[ht]
    \centering
    \includegraphics[width=0.5\linewidth]{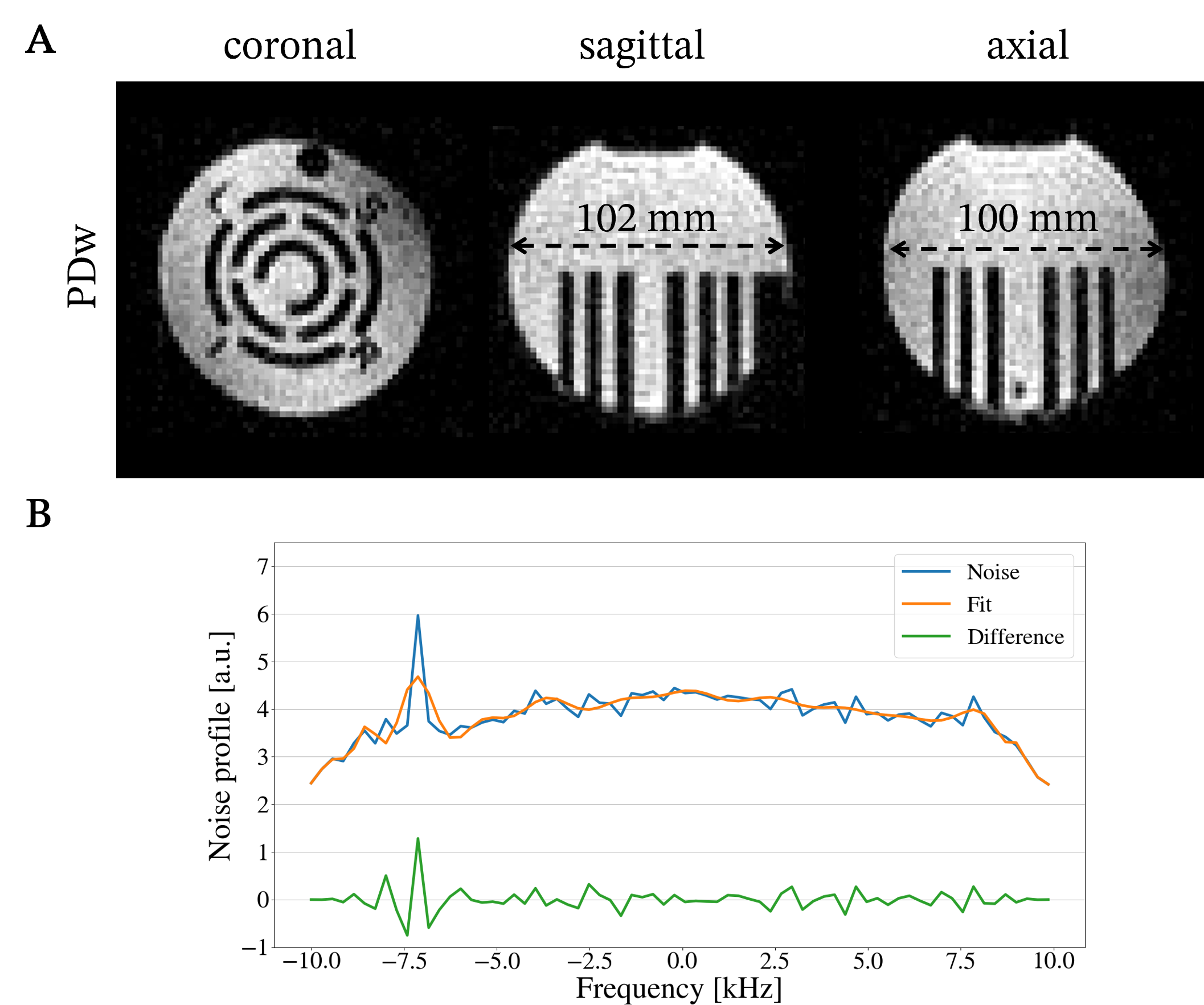}%
    \caption{MRI images of the \emph{Hello World} phantom with \gls{pdw} displayed in the coronal, sagittal, and axial orientations at the center of the phantom (A). The extent of the phantom in two dimensions is indicated by the dotted arrows. The images are rotated by \ang{45} degree in the coronal plane with respect to the scanner bore due to the rotation of the gradient axes. The frequency dependent noise profile (blue) from the noise acquisition together with a polynomial fit (orange) and their difference (green) shows a gentle slope with a noise spike at roughly \qty{-7.5}{\kilo\hertz} (B).}\label{fig:fig_qc}
\end{figure}

Figure~\ref{fig:fig_qc} shows \gls{pdw} images of the \emph{Hello World} phantom in three central planes. The printed features of the phantom including axis asymmetry, a reference point on the right side and the orientation letters were clearly visible. Note the \ang{45} rotation of the images in the coronal plane with respect to the scanner axes which is due to the rotation of the readout gradients as described earlier. Good geometric accuracy was achieved with \qty{2}{\milli\metre} deviation along the S-I direction and no deviation along the R-L direction with respect to the true dimensions of the phantom. Figure~\ref{fig:fig_qc}B shows the noise profile at different frequencies with a gentle slope and a noise spike at roughly \qty{-7.5}{\kilo\hertz}. Image SNR was 18.1 with a noise RMS value of \qty{27.4}{\micro\volt}. 

\subsection{Quantitative MRI}\label{results-qmri}

\begin{figure}[ht]
    \centering
    \includegraphics[width=0.5\linewidth]{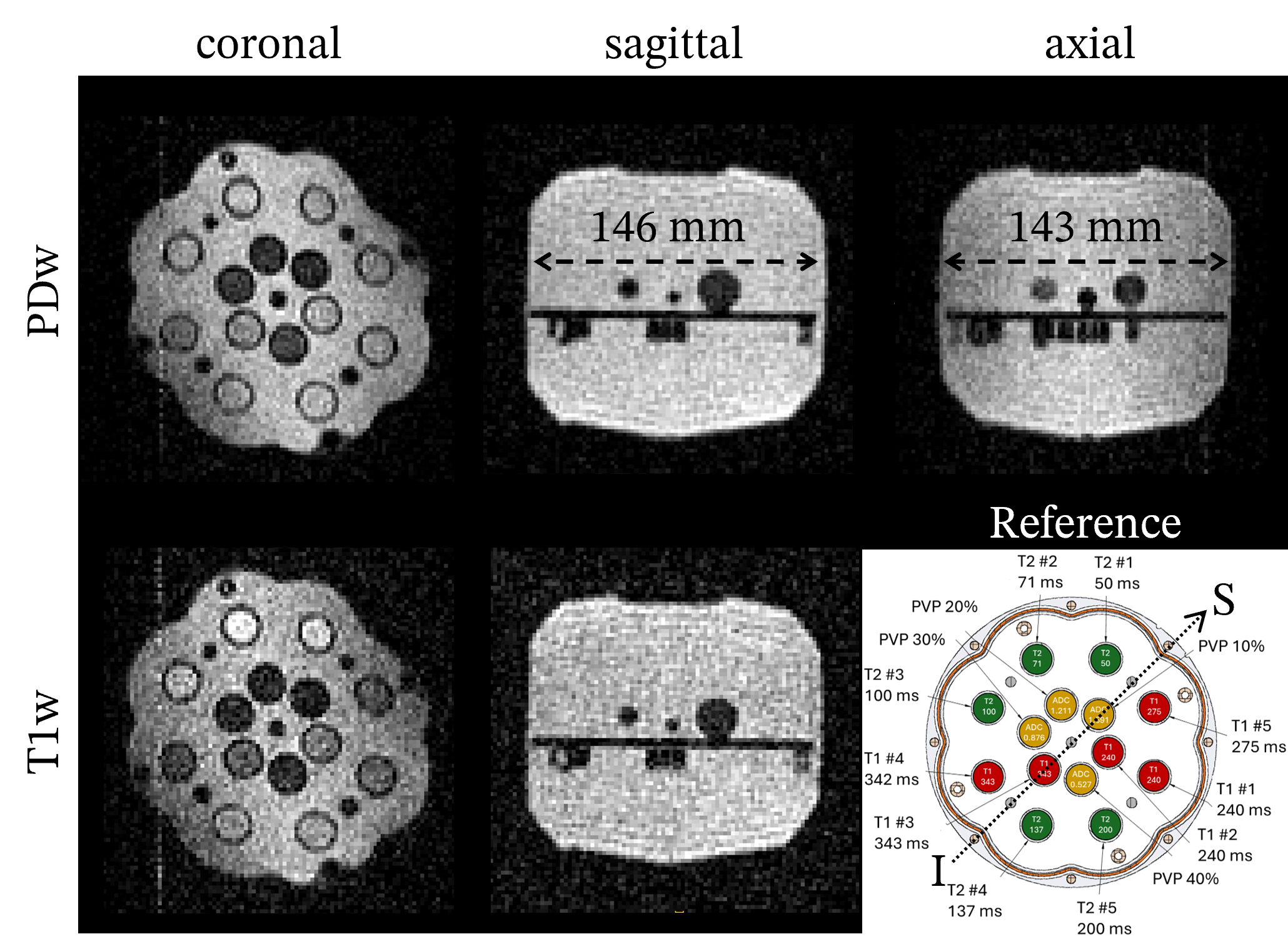}%
    \caption{MRI images of the \emph{CaliberMRI} phantom with \gls{pdw} (top) and \gls{t1w} (bottom) displayed in the coronal, sagittal, and axial orientations at a central section of the phantom. The various inclusions for \T1 and \T2 mapping are visible in the coronal slice, as illustrated in the coronal reference diagram (bottom right) which identifies the individual compartments and their nominal \T1, \T2, and \gls{adc} values with different percentages of Polyvinylpyrrolidone (PVP). The extent of the phantom in two dimensions is indicated by the dotted arrows. The images are rotated by \ang{45} degree in the coronal plane with respect to the scanner bore (dotted arrow along S-I direction) due to the rotation of the gradient axes.}\label{fig:fig_anat_caliber}
\end{figure}

\begin{figure}[ht]
    \centering
    \includegraphics[width=0.5\linewidth]{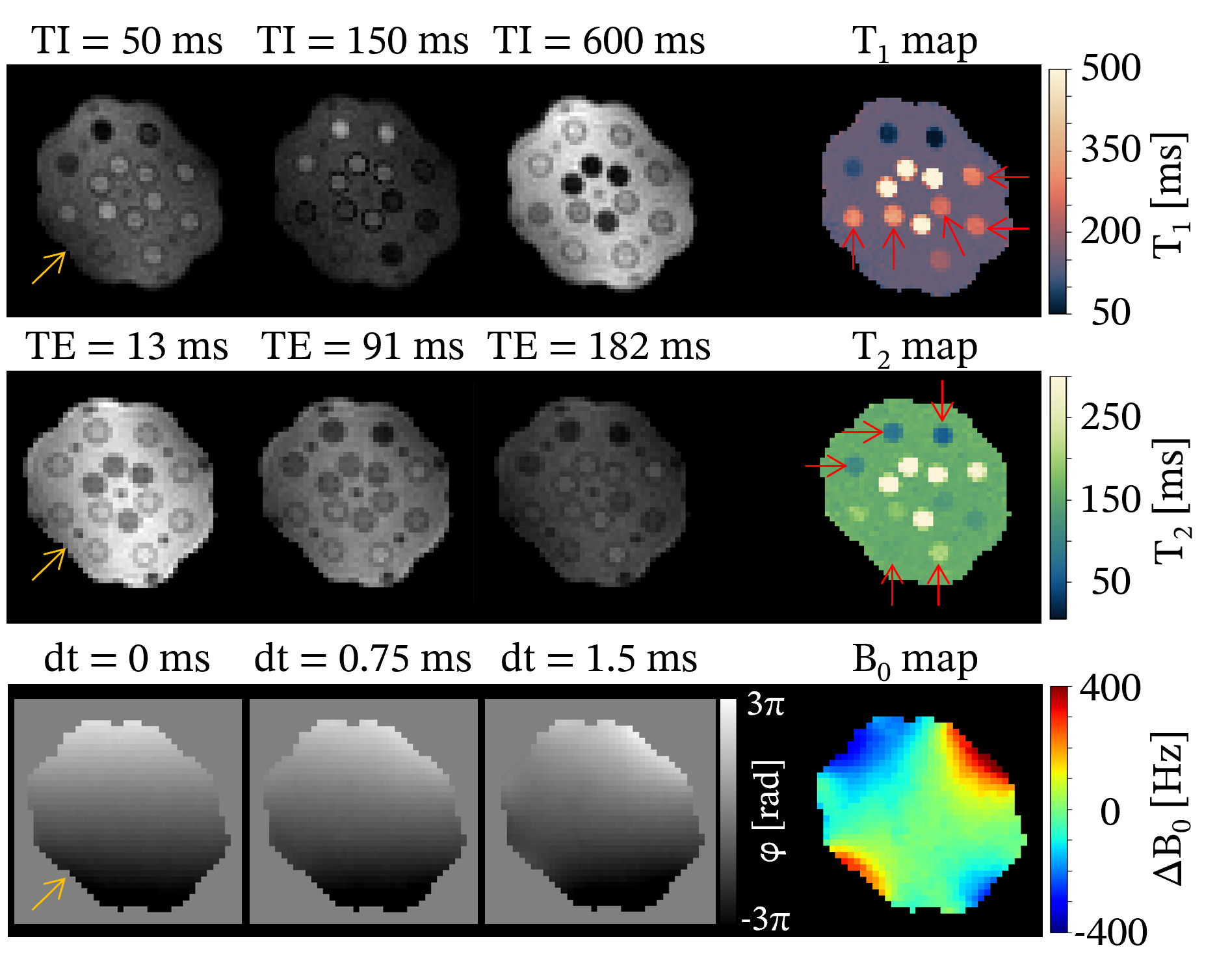}%
    \caption{MRI images and quantitative maps of a central coronal slice of the \emph{CaliberMRI} phantom from 3D sequences used for \T1 mapping (top), \T2 mapping (middle) and \B0 mapping (bottom). Top row: inversion recovery images at selected TI of \qtylist[list-units=single]{50;150;600}{\milli\second} used for \T1 mapping, together with the resulting \T1 map. Middle row: multi-echo images acquired at selected TE of \qtylist[list-units=single]{13;91;182}{\milli\second} used for \T2 mapping, together with the corresponding \T2 map. Bottom row: phase ($\phi$) images acquired at different echo-time offsets $\Delta t$ of \qtylist[list-units=single]{0;0.75;1.5}{\milli\second} for \B0 mapping, together with the resulting $\Delta B_0$ field map. Yellow arrows indicate regions of geometric distortions. Red arrows indicate the inclusions used for the comparison with the reference values provided by the phantom vendor.}\label{fig:fig_qmri}
\end{figure}

\begin{figure}[ht]
    \centering
    \includegraphics[width=0.5\linewidth]{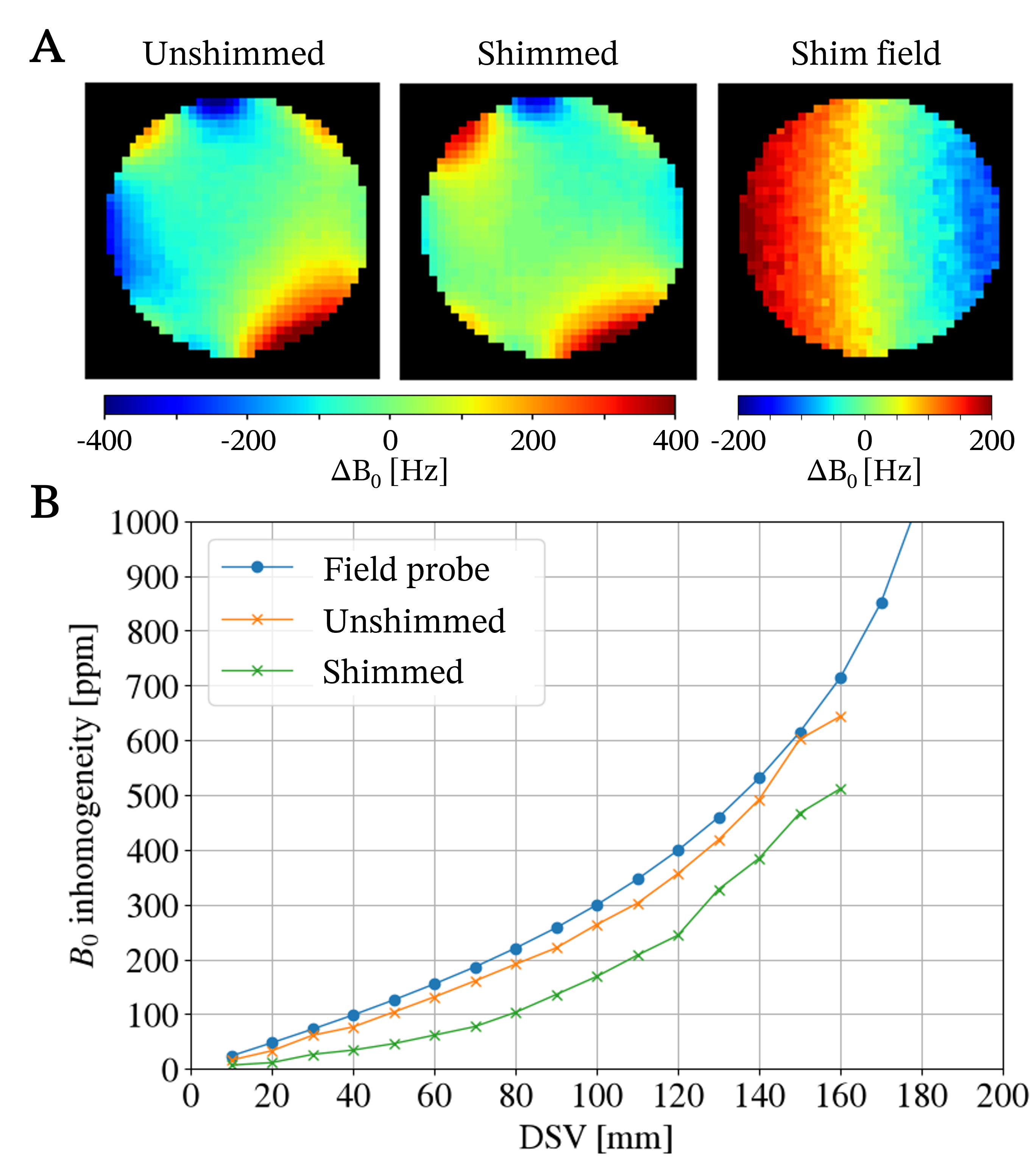}%
    \caption{Results of \B0 mapping using the spherical phantom. Maps of \B0 inhomogeneity with and without active shimming together with the shim fields derived by voxel-wise subtraction in a central axial slice (A). In addition, the \B0 inhomogeneity was calculated for increasing spherical volumes (DSV) in three configurations(B). The results based on the field probe (without active shimming) showed the highest inhomogeneity (blue), which were closely matched by the imaging based \B0 mapping without active shimming (orange). Active shimming markedly improved the homogeneity for all DSVs (green).}\label{fig:fig_b0}
\end{figure}

\begin{figure}[ht]
    \centering
    \includegraphics[width=0.5\linewidth]{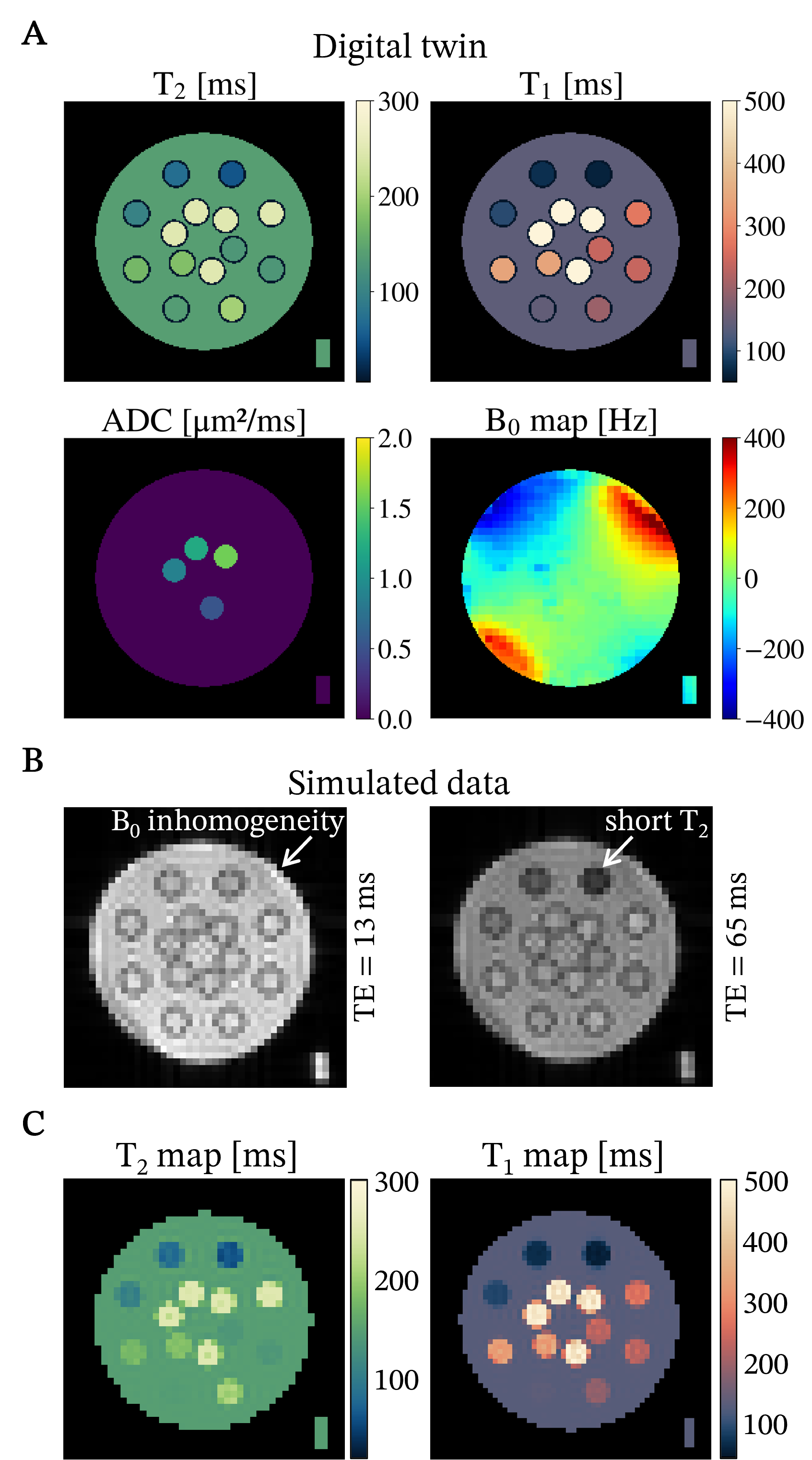}%
    \caption{Digital twin and simulation results for quantitative MRI validation. Digital twin of the phantom comprising spatially resolved \T2, \T1, \gls{adc}, and $\Delta B_0$ (A). Representative simulated magnitude images at \qty{13}{\milli\second} and \qty{65}{\milli\second}. Signal variations reflect the underlying relaxation properties of the inserts, while image distortions near the phantom boundary arise from the incorporated \B0 field inhomogeneity as indicated by the white arrows (B). Quantitative \T2 and \T1 maps reconstructed from the simulated data, demonstrating recovery of the prescribed relaxation properties of the digital phantom (C).}\label{fig:fig_sim}
\end{figure}

\begin{figure}[ht]
    \centering
    \includegraphics[width=\linewidth]{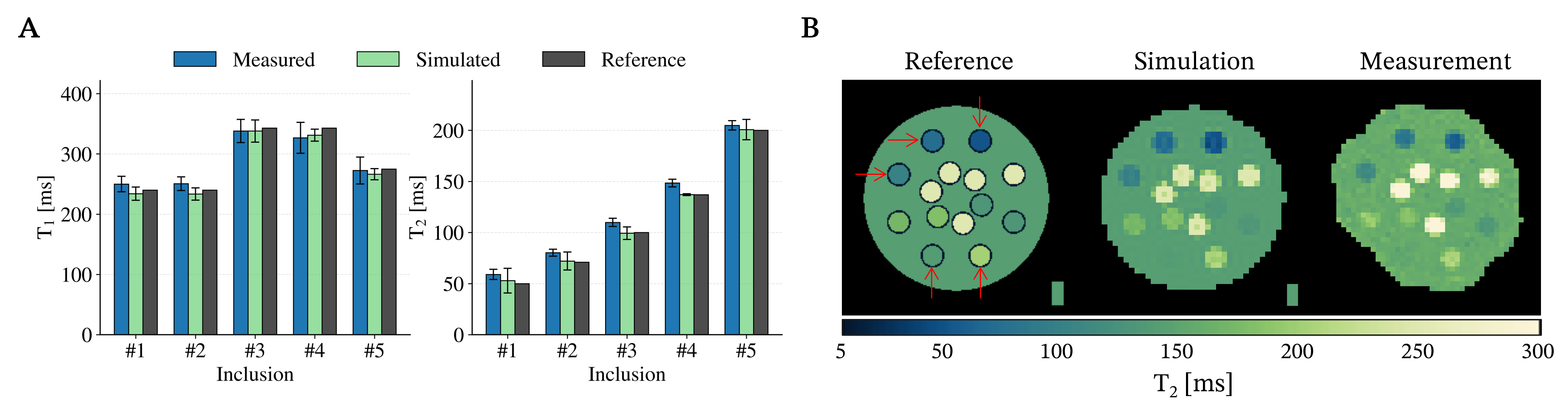}%
    \caption{Comparison of measured, simulated, and reference \T1 and \T2 relaxation times. Average values and standard deviations for five phantom inclusions for each contrast (A). Estimated \T1 (left) and \T2 (right) values obtained from MRI measurements (blue) and digital twin simulations (orange) are shown alongside the nominal reference values (gray). Error bars denote the standard deviation of the voxel-wise estimates within each inclusion. Visual comparison for \T2 mapping (B). Simulation, experiment and reference values showed very good agreement.}\label{fig:fig_barplots}
\end{figure}

MRI images with \gls{pdw} and \gls{t1w} of the \emph{CaliberMRI} phantom are displayed in Figure~\ref{fig:fig_anat_caliber} together with the reference diagram of the phantom. The scanner bore along the S-I line is indicated by the dotted arrow in the reference diagram. Key structural features could be easily identified from the images and the geometric proportions lied within the specified values. The images demonstrate the geometric integrity of the phantom and the visibility of all compartments across imaging planes and contrasts. SNR for \gls{pdw} was 16.8 and was reduced for \gls{t1w} images to 10.7 due to the reduced TR. A single \gls{emi} line is visible on the left side of the phantom. Again the images were rotated by \ang{45} degree in the coronal plane with respect to the scanner axes due to the rotation of the gradient axes. \gls{t1w} images correctly enhanced inclusions \T2\#1, \T2\#2 and \T2\#3 with short \T1 values of approximately \qty{50}{\milli\second}, \qty{70}{\milli\second} and \qty{100}{\milli\second}, respectively. Signal changes in \T1 inclusions were comparatively smaller, as all \T1 inclusions exhibited values above the selected TR for \T1-weighting.

Figure~\ref{fig:fig_qmri} shows the results for quantitative imaging of \T1 and \T2 relaxation times and \B0 mapping. Magnitude images at selected inversion and echo times are shown for \T1 and \T2 mapping, respectively. Phase images for \B0 mapping are shown for all time shifts.  The different inclusions for \T1 and \T2 mapping were visible in the magnitude images and the quantitative maps clearly differentiated the phantom compartments according to their relaxation properties, while the $\Delta B_0$ field map revealed the spatial distribution of magnetic field inhomogeneity across the phantom. The yellow arrows indicate regions were the phantom image was geometrically distorted.  Red arrows indicate selected compartments used for the comparison with the reference values provided by the phantom vendor. Most compartments showed distinct relaxation times relative to the surrounding background matrix. Measured \T1 values in the selected inclusion \#1 to \#5 were \qtylist[list-units=single]{250\pm13; 251\pm11; 338\pm19; 327\pm26; 273\pm22}{\milli\second}, respectively. \T2 values in the selected inclusion \#1 to \#5 were \qtylist[list-units=single]{59\pm5; 80\pm3; 110\pm4; 148\pm4; 205\pm5}{\milli\second}, respectively. 

Figure~\ref{fig:fig_b0} summarizes the results for \B0 mapping with the spherical phantom. In Figure~\ref{fig:fig_b0}A \B0 maps with and without active shimming are displayed together with the shim field determined by voxel-wise subtraction. Active shimming reduced the overall inhomogeneity of the \B0 field as illustrated by the removal of first-order terms of the unshimmed field, which is further illustrated by the estimated shim field. First-order shim gradients, predominantly along the A-P direction were observed. The shim field was further decomposed along the gradient axes to retrieve the underlying shim currents with values of \qtylist[list-units=single]{77; 60; -6}{\milli\volt}, which closely matched the actual shim currents employed by the console of \qtylist[list-units=single]{80; 59; 0}{\milli\volt} across the channels. In Figure~\ref{fig:fig_b0}B, the \B0 inhomogeneity for an increasing \gls{dsv} for three configurations is shown. Image based \B0 mapping without active shimming closely matched the findings from the field probe acquisition, but was on average \qty{12}{\percent} lower. With active shimming, the \B0 inhomogeneity was reduced by \qty{45}{\percent} with respect to the field probe averaged over the different \glspl{dsv}.  

The simulation results are presented in Figure~\ref{fig:fig_sim}. Figure~\ref{fig:fig_sim}A illustrates the digital phantom with specified values for \T2, \T1, \gls{adc}, \B0 inhomogeneity. In Figure~\ref{fig:fig_sim}B, simulation results after image reconstruction are presented for selected echo times of \qty{13}{\milli\second} and \qty{65}{\milli\second}, respectively. A pronounced signal decay was observed for the \T2\#1 inclusion with the shortest \T2 time. Moreover, signal alterations and a slight phantom deformation were observed, which disappeared in the absence of \B0 inhomogeneity during signal simulation. Figure~\ref{fig:fig_sim}C gives the result of the reconstructed \T2 and \T1 map based on the simulated data, which showed a good agreement with the input data. The framework enabled realistic simulation of low-field MRI experiments, including relaxation contrast and system-related field imperfections. \T2 values from simulated data in the selected inclusion \T2\#1 to \T2\#5 were \qtylist[list-units=single]{53\pm12; 72\pm9; 99\pm6; 137\pm1; 201\pm10}{\milli\second}, respectively. \T1 values from simulated data in the selected inclusion \T1\#1 to \T1\#5 were \qtylist[list-units=single]{234\pm11; 234\pm10; 338\pm18; 331\pm10; 266\pm9}{\milli\second}, respectively.

Figure~\ref{fig:fig_barplots} summarizes the results for \T1 and \T2 mapping from measurements, simulations and vendor specified values, showing mean and standard deviation for each inclusion. In addition, the relevant \T2 maps are shown on the right side. Overall very good agreement between the vendor specified values and the measurements was achieved. For \T1 mapping, inclusion \T1\#1 and \T1\#2 with short \T1 values were slightly overestimated with an error of \qty{4}{\percent}, whereas the inclusions with high \T1 values \T1\#2, \T1\#3 and \T1\#4 were underestimated with \qty{-1}{\percent}, \qty{-5}{\percent} and \qty{-1}{\percent}, respectively. \T2 mapping showed a consistent overestimation for all inclusions with errors of \qty{18}{\percent}, \qty{13}{\percent}, \qty{10}{\percent}, \qty{8}{\percent} and \qty{2}{\percent} for inclusion \T2\#1 to \T2\#5, respectively. Interestingly, this observation was not reproduced by the simulation as the results matched the reference values very well. In contrast to the \T1 mapping results, this ambiguity points towards a sequence independent issue for \T2 mapping.

\subsection{System Replication}\label{results-reproducibility}


\begin{figure}[ht]
    \centering
    \includegraphics[width=0.5\linewidth]{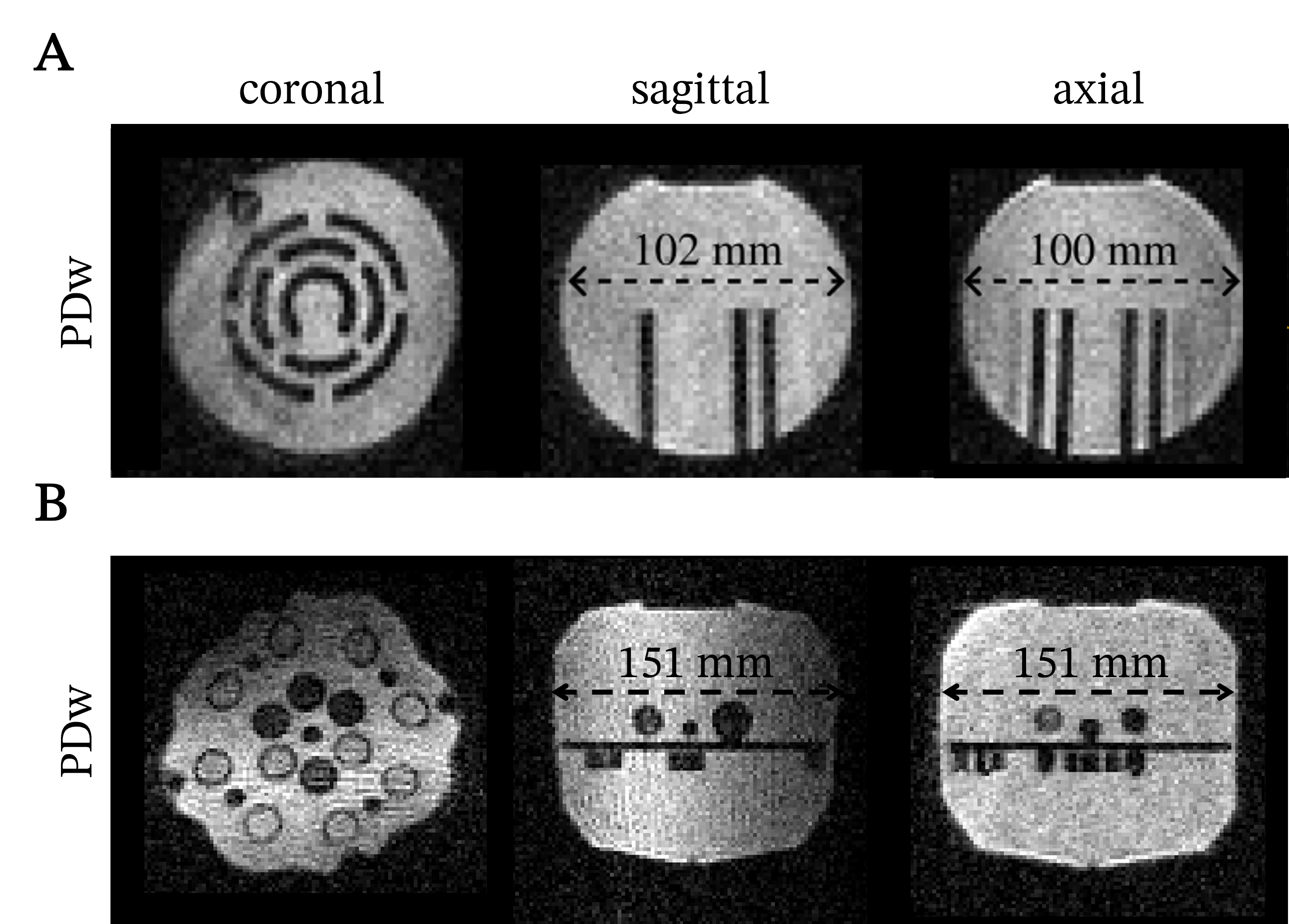}%
    \caption{Replication of MR imaging with different rebuilds of the reference scanner for structural imaging with \gls{pdw}.  MRI images of the \emph{Hello World} phantom acquired at TU Graz (A). MRI images of the \emph{CaliberMRI} phantom acquired at \gls{inrim} (B). Displayed are the coronal, sagittal, and axial orientations at the center of the respective phantom. The extend of the phantom in two dimensions is indicated by the dotted arrows.}\label{fig:fig_qc_graz}
\end{figure}

Figure~\ref{fig:fig_qc_graz} summarizes the image acquisition results from TU Graz and \gls{inrim}. Figure~\ref{fig:fig_qc_graz}A shows \gls{pdw} images of the \emph{Hello World} phantom in three central planes imaged with the TU Graz system configuration. Similar geometric accuracy compared to the reference scanner was achieved with \qty{2}{\milli\metre} deviation along the S-I direction and no deviation along the R-L direction. Image SNR was comparable to the reference implementation at 17.3 with a noise RMS value of \qty{13.3}{\micro\volt}. 
Figure~\ref{fig:fig_qc_graz}B shows \gls{pdw} images of the \emph{Caliber} phantom acquired with the \gls{inrim} system. The \qty{5}{\milli\meter} difference of the measured phantom width along the S-I and R-L directions suggests that the gradient coil sensitivities along these direction should be better calibrated. The measured SNR equaled \num{12.2} with an RMS noise voltage equal to \qty{18.9}{\micro\volt}.

\section{Discussion}\label{discussion}

In this work, we presented an end-to-end open source reference system for portable low-field MRI that combines openly documented hardware, scanner control software, calibration routines, imaging sequences and quantitative MRI results within a single reproducible framework. The system demonstrated robust imaging performance, quantitative \T1, \T2 and \B0 mapping capabilities, and standardized quality-control procedures based on dedicated phantoms. By providing open documentation, design files, and accompanying datasets, the proposed platform lowers the barrier to replicating and benchmarking low-field MRI systems across institutions. We anticipate that this reference implementation will support cross-site validation, facilitate collaborative development, and contribute to the standardization and broader adoption of portable low-field MRI technologies. The technological foundation was complemented by an open source documentation for regulatory purposes which, among others, include a list of applicable standards, risk analysis, technical dossier summary file and instructions for use. These blueprints are a valuable resource for a fast and resource effective transition of these and other prototypes into medical devices, ultimately saving costs for the healthcare system.\cite{winter2019, moritz2019oshvalue}

The user interaction with the scanner was provided by a web-based acquisition platform. Its asset-centric pipeline allows to streamline further processing steps, such as denoising or distortion correction. The modular and scalable architecture could be deployed on a single workstation, however, it also provides the flexibility to be offloaded to a dedicated compute node or to be distributed across a compute cluster to share computational resources across multiple systems and scale them as needed.

The noise characterization of the system revealed an elevated noise level of 1.4 when connecting the RF coil in contrast to 1.15 with a \qty{50}{\ohm} resistance termination. Guallart-Naval et al.\cite{guallart2026electromagnetic} reported very similar noise levels for the fully assembled system and recommended a noise level of \qtyrange{\sim1.2}{1.5} for imaging. Future work could address the remaining contributions of the different hardware components, such as the preamplifier noise figure, and residual \gls{emi}. Different \gls{emi} mitigation strategies already have been proposed which could be easily integrated into the existing setup.\cite{pfitzer2026, schote2025, zhaoRobustEMIElimination2024, yangActiveEMISuppression2022, srinivas2022}

One potential reason could be the choice of the thin \qty{7}{\micro\metre} copper RF shield used in the proposed setup. While it allowed for relatively short gradient-induced eddy currents, it affected the shielding effectiveness and the RF coil Q-factor. RF coil simulations modeled the shield with Surface Impedance Boundary Conditions. This allowed to model the losses on the copper shield without having to resort to a sub-millimeter mesh to model the skin effect, as long as the latter is much smaller than the thickness of the shield\cite{bib:yuferev1, zilberti1, bib:comsol_SIBC}. 
As expected form a copper shield thickness of about \qty{80}{\percent} of the skin depth at \qty{2}{\mega\hertz}, like the one presently used in the reference system, measurements showed higher RF coil losses with respect to those simulated. To retain the high Q-factor of the coil and obtain improved shielding with quickly vanishing eddy currents, a thicker shield and proper segmentation should be employed as proposed by de Vos et al.\cite{devosSegmentedRFShield2024}, who measured time constants between \qtyrange{50}{100}{\micro\second} for a \qty{75}{\micro\metre} segmented shield.

The eddy current characterization demonstrated highly consistent behavior across all gradient channels and polarities, with time constants between \qty{27}{\micro\second} and \qty{32}{\micro\second} and normalized fitting residuals below \qty{1}{\percent}. These findings indicate that the conductive structures introduced into the imaging volume, including the RF shield, did not introduce long-lived field perturbations that would possibly have degraded image quality and constrain pulse-sequence design. Nevertheless, advanced methodologies for eddy current compensations such as gradient waveform pre-emphasis\cite{stich2018} or compensation of concomitant field effects\cite{wang2025} may further improve image quality and are likely necessary for non-cartesian trajectories.

In addition to the calibration and characterization of the fully assembled system, we presented a reproducible setup to assess low-field MRI imaging performance using the \textit{OSI\textsuperscript{2}~ONE \emph{Hello World}} phantom. We believe this setup to be a good landmark in the development of an MRI system as it allows a comparative assessment of the imaging performance at an early stage before dedicated system improvements and sequence optimizations take place. During system development, repeated phantom measurements proved valuable for detecting unintended changes following hardware or software modifications and thus served as a practical tool for continuous quality assurance.

Another focus of this work was to use quantitative MRI for performance assessment using measurable physical parameters rather than image quality metrics alone. We performed multi-parametric mapping of the quantitative \emph{CaliberMRI} phantom using the presented hardware and software configuration. Structural scans were used to evaluate image quality, while quantitative \T1, \T2 and \B0 mapping provided insight into the accurate measurement of relaxation times and field inhomogeneities. In contrast to the images of the \emph{Hello World} phantom, geometric distortions were apparent using the \emph{CaliberMRI} phantom which were linked to \B0 inhomogeneity and gradient non-linearity at a larger FOV. Five inclusions each for \T1 and \T2 relaxation times were compared between imaging results, simulations and reference values. We found that \T1 inclusions were accurately mapped with an average absolute error of \qty{-3.1\pm1.8}{\percent} and \T2 inclusions were consistently overestimated by \qty{10.4\pm5.8}{\percent}. Simulations were very accurate for both \T1 mapping and \T2 mapping with average absolute deviations of \qty{2.7\pm0.8}{\percent} and \qty{1.7\pm2.4}{\percent}, respectively. Possible explanations for the \T2 overestimation might be imperfect refocusing pulses during the echo train causing a mismatch with the signal model or an elevated Rician noise floor mimicking a prolonged signal decay. In addition, signal alterations due to Gibbs ringing or partial volume effects might have affected the reconstruction due to the relatively large voxels used.

Image based \B0 mapping was successfully performed in two different phantoms with and without active shimming. Active shimming reduced the unshimmed \B0 homogeneity by \qty{37.5}{\percent}. Analysis of the spherical phantom showed that the employed shim fields can accurately be recovered by subtracting individual \B0 maps. Vice versa, the unshimmed \B0 map allows to directly derive the optimal shim fields given by the first order terms of a least-squares polynomial approximation. This method could robustly replace our initial approach of optimizing the shim current combinations using a series of \gls{fid} signals during the system calibration. Moreover, the known \B0 map could improve the image reconstruction pipeline and support the reconstruction of undistorted image data. We further included the acquired \B0 map in the simulation framework, and observed \B0 induced signal alteration at the edges of the phantom together with a slight image distortion along the readout direction. This may prove useful for more realistic MRI simulations, especially for low bandwidth acquisitions, strong \B0 inhomogeneities, or voxel-wise comparisons.

The experiments involving the replicated scanner configurations showcased the modularity of the system. Components such as the console, RF coils, \gls{rfpa}, and \gls{gpa} can be easily exchanged with other available options.\cite{Negnevitsky_2023} Comparing different system implementations can be beneficial, as small implementation errors can lead to considerable degradation in image quality. The system replication at TU Graz revealed similar imaging quality despite differences in the hardware components and pulse sequence used. Additionally, the system was successfully replicated at \gls{inrim}, yet, with lower image quality and visible image distortions. At the time of writing, the system has only recently been assembled and markedly improved results are expected in the near future when the imaging chain is fully optimized. Image SNR at \gls{inrim} was \qty{27}{\percent} lower than the reference, but the measured noise levels were comparable.

Despite our effort to present a fully functional and performative reference system for portable low-field, further work is needed to continuously improve the system. During the system setup, frequent tuning and matching of the RF coil to different loads and resonance frequencies (changes with ambient temperature) added an extra preparatory step which could be omitted if an automatic tuning and matching network would be in place.\cite{algarin2024marge}

A further limitation arises from the rotated gradient coordinate system used to achieve favorable coil efficiencies. Although this design improves gradient performance, it complicates sequence implementation and image orientation. In practice, deviations from perfectly orthogonal gradient fields prevented straightforward correction through simple coordinate transformations. Future work should therefore include full gradient field calibration and image reconstruction methods that account for the measured gradient field geometry. Such approaches could eliminate the residual image rotation and improve geometric fidelity. 

Finally, the quantitative imaging framework for the \emph{CaliberMRI} phantom can be further expanded, as the phantom offers structural landmarks for further quality control procedures. So far, distortion fiducial, resolution inserts, and inclusions with defined diffusivity, remained unexplored. Incorporating these measurements would further strengthen the platform as a common benchmark for low-field MRI systems. Alignment with emerging initiatives such as the \emph{UNITY} network\cite{ABATE2024101397} may additionally support harmonized quality assurance procedures across institutions and facilitate broader adoption of standardized evaluation methods. 

Overall, the proposed platform demonstrates that reproducible, openly documented, and quantitatively characterized low-field MRI systems are feasible using predominantly open source hardware and software components. By combining scanner hardware, calibration procedures, quality-control phantoms, quantitative MRI methods, and simulation tools within a single ecosystem, the presented reference system provides a foundation for reproducible research and collaborative development in portable low-field MRI. The presented configuration with all hardware components used in this work can be potentially assembled towards a clinical system as illustrated in \ref{fig:system-architecture}. Therefore we provided additional documentation regarding safety testing tailored to portable MRI systems which can serve as blueprints for the conformity assessment process according to the \gls{mdr} (EU) 2017/74\cite{eu_mdr_2017_745}.

\section{Conclusion}\label{conclusion}

We presented an end-to-end open source reference system for portable low-field MRI that integrates openly documented hardware, scanner control software, system calibration and characterization, image reconstruction, and quantitative image analysis into a reproducible imaging platform. All major hardware components, software repositories, calibration routines, characterization procedures, and imaging datasets are made available under open licenses, enabling independent reproduction and verification by other groups. Together with standardized phantoms and measurement procedures, this establishes a practical foundation for cross-site benchmarking and methodological comparison in low-field MRI. Additionally, the documentation blueprints can be used as a basis towards approval for clinical in-vivo studies. We hope this work sparks further collaboration and open source developments, which can be shared with the community and translated into affordable clinical applications. 

\section*{Contributions}

The following \emph{CRediT} (Contribution Roles Taxonomy) applies:
\textbf{David Schote:} conceptualization, methodology, software, validation, formal analysis, investigation, resources, data curation, writing – original draft, visualization, project administration.
\textbf{Helge Herthum:} conceptualization, methodology, software, validation, formal analysis, investigation, resources, data curation, writing – original draft, visualization, project administration.
\textbf{Umberto Zanovello:} hardware, visualization, methodology, writing – review and editing
\textbf{Julia Pfitzer:} visualization, methodology, writing – review and editing
\textbf{Ivo Jutte:} visualization, methodology, writing – review and editing
\textbf{Jan Gregor Frintz:} hardware, methodology, validation, writing – review and editing
\textbf{Ilia Kulikov:} hardware, methodology, software, writing – review and editing
\textbf{Sebastian Littin:} hardware, methodology, writing – review and editing
\textbf{Frank Seifert:} resources, supervision, writing – review and editing
\textbf{Christoph Kolbitsch:} methodology, supervision, writing – review and editing, funding acquisition
\textbf{Lukas Winter:} conceptualization, methodology, resources, writing – review and editing, supervision, project administration, funding acquisition

\section*{Acknowledgments}

The authors would like to thank Danny de Gans for designing and documenting the custom-built \gls{gpa} used in this work; Sebastian Schachel and Christian Engler for designing the \textit{OSI\textsuperscript{2}~ONE \emph{Hello World}} phantom and corresponding railing system; Christian Meumann, Reiner Montag and Tobias Mohr for the integration of electronics, manufacturing of the gradient coils, RF shield, RF coil and 3D printing further pieces; Johannes Behrens, Christoph Dinh and Ben Wilhelm-Feldbusch for the support in the implementation of ScanHub; Tom O'Reilly for contributing to the Nexus software code and to the magnet design. Furthermore, the authors would like to thank CMI Medical for their very valuable contribution on the MDR documentation.

\section*{Conflicts of Interest}

The authors declare no potential conflicts of interest.

\section*{Data Availability Statement}

Hardware documentation is open source and available at \url{https://gitlab.com/osii}. 
The data required to reproduce the results are freely available via Zenodo (\url{doi.org/10.5281/zenodo.21807140}). We provide Jupyter notebooks for noise characterization, \T1 mapping, \T2 mapping, and MRI simulations demonstrating the functionality in more detail.\cite{jupyter_notebooks}

\bibliography{bibliography}
\bibliographystyle{ieeetr}

\end{document}